\documentclass[12pt]{article}
\usepackage{epsfig}
\usepackage{cite}
\usepackage{amsmath}
\usepackage{hhline}
\usepackage{amssymb}
\usepackage{times}

\newlength{\dinwidth}
\newlength{\dinmargin}
\def\Journal#1#2#3#4{{#1} {\bf #2} (#3) #4}

\def\ar#1#2#3   {\Journal{\em Ann. Rev. Nucl. Part. Sci.}{\bf#1}{#2}{#3}}
\def\err#1#2#3  {\Journal{\em Erratum}{\bf#1}{#2}{#3}}
\def\ib#1#2#3   {\Journal{ibid.}{\bf#1}{#2}{#3}}
\def\ijmp#1#2#3 {\Journal{\em Int. J. Mod. Phys.}{\bf#1}{#2}{#3}}
\def\jetp#1#2#3 {\Journal{\em JETP Lett.}{\bf#1}{#2}{#3}}
\def\mpl#1#2#3  {\Journal{\em Mod. Phys. Lett.}{\bf#1}{#2}{#3}}
\def\nim#1#2#3  {\Journal{\em Nucl. Instrum. Meth.}{\bf#1}{#2}{#3}}
\def\nc#1#2#3   {\Journal{\em Nuovo Cim.}{\bf#1}{#2}{#3}}
\def\np#1#2#3   {\Journal{\em Nucl. Phys.}{\bf#1}{#2}{#3}}
\def\pl#1#2#3   {\Journal{\em Phys. Lett.}{\bf#1}{#2}{#3}}
\def\prep#1#2#3 {\Journal{\em Phys. Rep.}{\bf#1}{#2}{#3}}
\def\prev#1#2#3 {\Journal{\em Phys. Rev.}{\bf#1}{#2}{#3}}
\def\prl#1#2#3  {\Journal{\em Phys. Rev. Lett.}{\bf#1}{#2}{#3}}
\def\ptp#1#2#3  {\Journal{\em Prog. Th. Phys.}{\bf#1}{#2}{#3}}
\def\rmp#1#2#3  {\Journal{\em Rev. Mod. Phys.}{\bf#1}{#2}{#3}}
\def\rpp#1#2#3  {\Journal{\em Rep. Prog. Phys.}{\bf#1}{#2}{#3}}
\def\sjnp#1#2#3 {\Journal{\em Sov. J. Nucl. Phys.}{\bf#1}{#2}{#3}}
\def\spj#1#2#3  {\Journal{\em Sov. Phys. JEPT}{\bf#1}{#2}{#3}}
\def\zp#1#2#3   {\Journal{\em Z. Phys.}{\bf#1}{#2}{#3}}
\def\ejc#1#2#3  {\Journal{\em Eur. Phys. J.}{\bf#1}{#2}{#3}}
\def\jetp2#1#2#3 {\Journal{\em J. Exp. Theor. Phys.}{\bf#1}{#2}{#3}}
\def\cpc#1#2#3 {\Journal{\em Comput. Phys. Commun.}{\bf#1}{#2}{#3}}

\def\gsim{\ \,\lower.25ex\hbox{$\scriptstyle\sim$}\kern-1.30ex%
\raise 0.55ex\hbox{$\scriptstyle >$}\ \,}
\def\lsim{\ \,\lower.25ex\hbox{$\scriptstyle\sim$}\kern-1.30ex%
\raise 0.55ex\hbox{$\scriptstyle <$}\ \,}
\newcommand{\rb}[1]{\raisebox{1.5ex}[-1.5ex]{#1}}
\newcommand{\bdec}{$b\ra\jpsiw+X$}
\newcommand{\bcon}{$b$ contribution}
\newcommand{\ccbar}{\ensuremath{\mit c\overline{c}}}
\newcommand{\psits}{$\psi(2S)$}
\newcommand{\cms}{center of mass}

\newcommand{\dsdz}{$d\sigma/dz$}
\newcommand{\dsdptt}{$d\sigma/d\ptt$}
\newcommand{\ppbar}{$p\overline{p}$}
\newcommand{\bbbar}{b\overline{b}}

\newcommand{\ra}{\rightarrow}
\newcommand{\picb}{\mbox{pb}^{-1}}
\newcommand{\gev}{\,\mbox{\rm GeV}}
\newcommand{\mev}{\,\mbox{\rm MeV}}
\newcommand{\GeV}{\,\mbox{\rm GeV}}

\newcommand{\gevt}{\,\mbox{\GeV$^2$}}

\newcommand{\qsq}{\ensuremath{Q^2} }
\newcommand{\ptt}{\ensuremath{p_{t,\psi}^2}}
\newcommand{\ptr}{\ensuremath{p_t} }
\newcommand{\wgp}{\ensuremath{W_{\gamma p}}}
\newcommand{\sgp}{\ensuremath{\sigma_{\gamma p}}}
\newcommand{\jpsi}{$J/\psi$}
\newcommand{\mpsit}{$M_{J/\psi}$}
\newcommand{\psiprime}{$\psi'$}
\newcommand{\jpsiw}{J/\psi}
\newcommand{\wgpt}{\ensuremath{W^2_{\gamma\,p}}}

\newcommand{\ptpsi}{\ensuremath{p_{t,\psi}}}
\newcommand{\csm}{Colour Singlet Model}
\newcommand{\colsing}{colour singlet}
\newcommand{\coloct}{colour octet}
\begin{document}

\begin{titlepage}

\begin{flushleft}
\noindent
%Date:   \today; \the\time            \\
%Version: 4.0          \\
%Editors: K. Kr\"uger, B. Naroska           \\
%Referees: C. Grab, F. Sefkow          \\
%Comments by  24.03. 2002         

DESY-02-059 \hfill ISSN 0418-9833\\
May 2002
\end{flushleft}
 
\vspace*{3cm}

\begin{center}
\begin{Large}

{\bf Inelastic Photoproduction of \boldmath$J/\psi$ Mesons at HERA}

\vspace{2cm}

H1 Collaboration

\end{Large}
\end{center}

\vspace{2cm}

\begin {abstract}
\noindent
An analysis of inelastic photoproduction of \jpsi\ mesons is presented
using data collected at the $ep$ collider HERA corresponding to an integrated 
luminosity of above $80\,\picb$. 
 Differential and double differential cross sections  are measured in a wide 
kinematic region: $60<W_{\gamma p}<260\,{\rm GeV}$, $1<\ptt< 60\,\gevt$ and
$0.05<z<0.9$, where $z$ is the fraction of the energy of the 
exchanged photon transferred to the \jpsi\ meson in the rest frame of the
target proton.
Cross sections at $z\lsim0.3$ are presented for the first time.
%Theoretical calculations, which are available within the \csm\ at NLO for direct photon 
%processes, give a good description of the data in the  medium $z$  region ($0.3<z<0.9$) and  
%up to the highest \ptt\ values.
Theoretical calculations within the \csm\ at NLO for direct photon 
processes are shown to give a good description of the data in the medium $z$  region ($0.3<z<0.9$)  
up to the highest \ptt\ values.
A calculation using a  $k_t$ factorisation approach
in LO in the \csm\ is also able to describe these data.
 The data in the full $z$ range are also compared to LO calculations within a non-relativistic 
QCD framework including \coloct\ and \colsing\ contributions for 
direct and resolved photons. It seems
possible to reconcile data and theory with modest contributions from
\coloct\ processes. 
 The polarisation of the \jpsi\ meson is measured as a function of $z$ and 
\ptpsi\ and is reasonably described by the theoretical predictions.
\end{abstract}
\vspace{1.5cm}

\begin{center}
To be submitted to Eur.~Phys.~J.~C
\end{center}
\vspace{1.5cm}
\end{titlepage}
%-------------------------------------------------------------------------------

\begin{flushleft}
%-- H1AUTS Author list by names 
%-- Status: Tue Dec 18 09:41:12 MET 2001  Number of authors = 330 

C.~Adloff$^{33}$,              %WUPP-LEFT      07/01           Adloff              
V.~Andreev$^{24}$,             %LPI -PD        8/88            Andreev             
B.~Andrieu$^{27}$,             %ECPL-LEFT      09/01           Andrieu             
T.~Anthonis$^{4}$,             %ANTW-ST        11/99           Anthonis            
A.~Astvatsatourov$^{35}$,      %ZEUT-ST        02/98           Astvatsatourov      
A.~Babaev$^{23}$,              %ITEP-PD        8/88            Babaev              
J.~B\"ahr$^{35}$,              %ZEUT-PD        8/88            Baehr               
P.~Baranov$^{24}$,             %LPI -PD        8/88            Baranovp            
E.~Barrelet$^{28}$,            %PARI-PD        11/99           Barrelet            
W.~Bartel$^{10}$
% ...

%-- H1 Institutes 
\bigskip{\it
% ...
$ ^{4}$ Inter-University Institute for High Energies ULB-VUB, Brussels;
          Universiteit Antwerpen (UIA), Antwerpen; Belgium$^{ c}$ \\
% ...
$ ^{10}$ DESY, Hamburg, Germany \\
% ...
$ ^{23}$ Institute for Theoretical and Experimental Physics,
          Moscow, Russia$^{ l}$ \\
 $ ^{24}$ Lebedev Physical Institute, Moscow, Russia$^{ e}$ \\
% ...
$ ^{27}$ LPNHE, Ecole Polytechnique, IN2P3-CNRS, Palaiseau, France \\
 $ ^{28}$ LPNHE, Universit\'{e}s Paris VI and VII, IN2P3-CNRS,
          Paris, France \\
% ...
$ ^{33}$ Fachbereich Physik, Bergische Universit\"at Gesamthochschule
          Wuppertal, Wuppertal, Germany \\
% ...
$ ^{35}$ DESY, Zeuthen, Germany \\
% ...
$ ^c$ Supported by FNRS-FWO-Vlaanderen, IISN-IIKW and IWT \\
% ...
$ ^e$ Supported by the Deutsche Forschungsgemeinschaft \\
% ...
$ ^l$ Partially Supported by Russian Foundation
      for Basic Research, grant    no. 00-15-96584 \\
}
\end{flushleft}

\newpage
%---------------------------------------------
\section{Introduction}
\begin{figure}[htbp]
\begin{center}
\unitlength1.0cm
\begin{picture}(15.5,3.8)   
   \put(-1.2,-1.){\epsfig{file=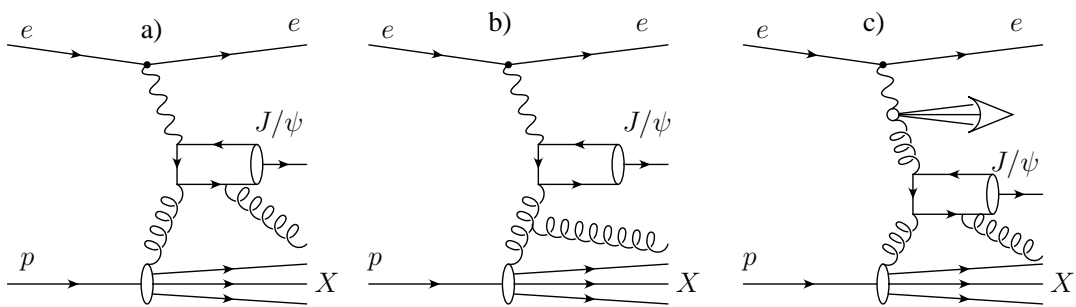,width=\textwidth,clip=}}
\end{picture}
\caption{\it Generic Feynman diagrams for inelastic \jpsi\ production.
a,b) direct photon processes; c) resolved photon process. In diagrams a) and c) the \ccbar\ pair 
leading to the formation of the \jpsi\ can be in a \colsing\ or octet state while in b) 
it can only be in a \coloct\ state. Additional soft gluons emitted during the hadronisation 
process are not shown.}
\label{fig_kine}
\end{center}
\end{figure}

The production of \jpsi\ mesons,  $ep\ra e\,\jpsiw\, X$, has been studied intensively 
 at HERA~\cite{susanne,Aid:1996dn,zeus97_jpsi,heradif}. The high available energy
 allows  the contributing mechanisms to be studied in a wide kinematic range in both \qsq\ 
and \wgp. Here \qsq\ is the negative squared four-momentum 
 of the exchanged photon and \wgp\ is the centre of mass energy of
the photon proton system.
In this paper we concentrate on the inelastic process where,  
in the proton rest system, the \jpsi\ meson  carries a fraction 
$z\leq 0.9$ of the photon energy in contrast to 
diffractive processes where this fraction is close to 1.
The inelastic process is dominated by boson gluon fusion: a photon 
emitted from the incoming lepton interacts with a gluon from the 
proton to produce a \ccbar\ pair which subsequently forms the \jpsi\ meson.
Here we will consider only interactions of quasi-real photons%
\footnote{The regime 
of virtual photons is studied in \cite{susanne}.}, $\qsq\simeq0$. 
In this case the photon can  either couple to the $c$ quark directly 
(``direct'' processes, Fig.~\ref{fig_kine}a or b) 
or can interact via its hadronic component (``resolved'' processes, 
Fig.~\ref{fig_kine}c).
\\
Many models have been suggested to describe inelastic \jpsi\ production in 
the framework of perturbative Quantum Chromodynamics (pQCD) such as the 
\csm\ (CSM)\cite{colsing}, 
the colour evaporation model \cite{evap} and soft colour interactions \cite{soft}.
Recently, discussions have focussed on calculations based on a factorisation 
theorem in a non-relativistic QCD approach (NRQCD)\cite{nrqcd}. 
These theoretical descriptions differ in many details. One important difference concerns 
the states they allow for the intermediate \ccbar\ pair\ produced in
the hard boson gluon interaction in terms of  
angular momentum and colour. Furthermore they differ in the  non-perturbative 
description of the transition from the intermediate \ccbar\ pair to the 
\jpsi\ meson. 

%Recently, discussions have focussed on calculations based on the NRQCD 
%approach, in which contributions from \ccbar\ pairs in \coloct\ states are 
%predicted in addition to  \colsing\ contributions. 
%In contrast, 
In the CSM it is assumed that the intermediate 
\ccbar\ pair is only produced in the quantum state of the \jpsi\
meson, i.e. in a \colsing\ state with spin 1 and no orbital angular momentum. 
However, the  CSM predicts prompt production  rates of \jpsi\
and \psiprime\ mesons in \ppbar\ interactions, which are lower  by more than an order of 
magnitude than the rates seen by the CDF collaboration~\cite{cdf}. 
The NRQCD approach, in which contributions from \ccbar\ pairs in \coloct\ states are 
predicted in addition to \colsing\ contributions, can accommodate 
the measured \ppbar\ cross sections.
Long distance matrix elements (LDMEs) are used to 
describe the transition of the various
intermediate \ccbar\ states to  the \jpsi\ meson. 
%They can be extracted from  fits to the CDF data.  
In previous publications~\cite{Aid:1996dn,zeus97_jpsi} it was shown
that in photoproduction at HERA the \coloct\ contributions 
are not observed with the  magnitude predicted
by leading order NRQCD calculations, in which  
LDMEs are used which are extracted, also in leading order (LO), from \ppbar\ data.
However the data were well described by the
next-to-leading order (NLO) calculations in the CSM.
Subsequently theoretical efforts were made to improve the simultaneous
description 
of the photoproduction and \ppbar\ processes within NRQCD, e.g. the effects of higher order 
corrections have been estimated in the extraction of the LDMEs as well as in the 
photoproduction amplitudes \cite{sanchis, martin, wolf,kniehl} . 

Since our first publication~\cite{Aid:1996dn} the amount of data 
taken by the H1 experiment 
has increased and allows more detailed investigations in a larger 
region of phase space. Here we present an analysis
 in an extended  $z$ region, $0.05<z<0.9$ covering $60<\wgp<260\,\gev$ and
 transverse momenta of the \jpsi, $1\leq\ptt\leq 60\,\gevt$. In the
medium $z$ region, which has been analysed previously, mainly direct 
photon processes 
are expected. At the lowest $z$ values resolved processes are expected
to contribute to the cross section in addition.     

The data in the medium $z$ range are compared with two sets of calculations 
within the CSM, namely a NLO calculation \cite{csm_nlo} and a LO  Monte Carlo 
computation \cite{cascade, Saleev:1994fg} using a 
``$k_t$ factorisation'' approach\footnote{Here ``$k_t$'' refers to 
a transverse momentum component of the gluon entering the hard 
interaction.}~\cite{kt3,kt4}.  
A comparison of the data with LO calculations in the NRQCD~framework including 
\coloct\ contributions is carried out over the full range $0.05\leq z\leq0.9$. 
The decay angular distribution of the \jpsi\ meson is analysed yielding 
information on the polarisation of the \jpsi\ meson and the results are  
compared with predictions in the NRQCD~framework both with and without 
\coloct\ contributions \cite{beneke} and in a $k_t$ factorisation approach~\cite{baranov}.

\section{Theoretical Models for \boldmath\jpsi\ Production}
The salient features of the relevant theoretical calculations 
%which are compared to the data 
are briefly summarised here. A comprehensive overview can be found 
in~\cite{kraemer}.

 In the \csm\ \cite{colsing} for direct photon processes the \jpsi\ meson is 
produced via $\mbox{$\gamma g\ra c\overline{c}\,[1,^3\!S_1]+g$}$, where the quantum 
state of the  \ccbar\ pair is described in spectroscopic notation% 
\footnote{
 Spectroscopic notation: $^{2S+1}\!L_J$ where $S$, $L$ and $J$ denote 
 the spin, orbital and total angular momenta of the $c\bar{c}$ system that is 
 produced in the hard process. The first number in the square bracket 
indicates the colour state of the $\ccbar$ pair.}.
The transition from the \ccbar\ pair to the \jpsi\ meson is calculated in 
potential models and can be related to the measured leptonic decay width of 
the \jpsi\ meson. % \cite{colsing}.
For photoproduction, full NLO calculations, 
i.e. up to order $\alpha_s^3$, where $\alpha_s$  denotes the strong coupling, 
are available for the direct 
process~\cite{csm_nlo}. The NLO contributions lead to an 
increase of the cross sections by  roughly a factor 2, dependent on the
 transverse momentum of the \jpsi\ meson.
The resolved photon process is expected to be dominated by 
$\mbox{$g g\ra  c\overline{c}\,[1,^3\!S_1]+g$}$, for which only 
LO calculations are available.

These CSM calculations are performed using standard collinear gluon density
distributions. An alternative $k_t$ factorisation approach~\cite{kt3,kt4} 
has recently been successfully applied to the description of a variety of 
processes\cite{cascade}. In this approach the \jpsi\ production 
process is factorised into a $k_t$ dependent gluon density and a 
matrix element for off-shell partons. A LO  calculation within this approach 
is implemented in the Monte Carlo generator CASCADE~\cite{cascade}.

In the NRQCD approach the \jpsi\ production amplitude factorises into  short
distance terms for the creation  of  colour octet or singlet \ccbar\ states with
definite angular momenta and long distance terms
describing the transition of these \ccbar\ states to the \jpsi\ meson. 
A double expansion in the strong coupling parameter $\alpha_s$ and $v$,  
the relative velocity of quark and antiquark, is obtained.
In specific calculations \cite{kraemer,kniehl,wolf} only  
the most important contributions are kept. In this expansion the lowest order 
term in $v$ is the \colsing\ term. Assuming that all other terms 
do not contribute, the CSM is recovered. 

Whereas the short distance amplitudes are calculable in pQCD the 
LDMEs must be determined from experimental data at present. They are assumed to be
process independent, i.e. they can be determined for example in \ppbar\
collisions or from $B$ decays and can then be used in predictions for 
photoproduction.
Calculations for photoproduction based on the NRQCD approach are available 
in LO taking into account the contributions
$ [1,^3\!S_1],\ [8,^3\!S_1],\ [8,^1\!S_0]\ \mbox{and}\ [8,^3\!P_{J=0,1,2}]$.
The relative strength of the \coloct\ contributions depends 
crucially on the size of the corresponding LDMEs. Unfortunately the values
for the LDMEs important at HERA still show large uncertainties 
(a summary can be found in \cite{kraemer}). 
Early extractions of the LDMEs from \ppbar\ data were performed in LO. 
Subsequent estimates of higher order effects 
led to considerably smaller values~\cite{sanchis, martin, wolf,kniehl}.   

The first comparisons of NRQCD calculations with photoproduction
data ~\cite{Aid:1996dn,zeus97_jpsi} revealed large discrepancies between
data and theory at high values of $z$, the theory being a 
factor 3-5 higher than the data. 
In later calculations resumming of soft gluon emissions within the NRQCD 
expansion \cite{wolf} was shown to damp the \coloct\ contributions at high $z$.
The region of validity of these calculations is restricted to 
\jpsi\ production with high transverse momentum, a region where experimental data 
suffer from limited statistics.

All theoretical calculations contain a number of free parameters, e.g. the parton 
density distributions, the values of $\alpha_s$ and the charm quark mass $m_c$ 
as well as the choice of the renormalisation and factorisation scales. 
In the NRQCD approach the values of the octet LDMEs are additional parameters.
The comparison with the data in the NRQCD approach also suffers from the 
uncertainties associated with LO calculations. 
Although the NLO terms have not been calculated in the NRQCD  approach 
similar effects as in the CSM may be expected, where the NLO terms lead to an 
increase of the cross section of typically a factor $2$ with a 
strong \ptpsi\ dependence.
Measurements of the polarisation of the \jpsi\ meson should help to
discriminate between different theoretical descriptions, independently of 
normalisation uncertainties.   

Table~\ref{tab:model1} gives an overview of the calculations available in 
the low and medium $z$ regions.

\begin{table}[htb]
\centering
\begin{tabular}{|l|l|l|}
\hline
$z$ region & Contributing Process & Available theoretical calculations\\
\hline
 medium $z$ & direct photon process &  CSM NLO and LO, NRQCD LO\\
 ($0.3<z<0.9$)&&$k_t$ factorisation \\
low $z$& direct + resolved photon processes& CSM LO, NRQCD LO\\
 ($0.05<z<0.45$)&&\\
\hline
\end{tabular}
\caption{\it Overview of theoretical calculations compared with the present
data in the medium and low $z$ regions.}
  \label{tab:model1}
\end{table}

\section{Experimental Conditions}

The data presented here were collected in the years 1996--2000 and 
correspond to a total integrated luminosity of 
%$(82.8 \pm 1.2)\mbox{~pb}^{-1}$  and 
$(87.5 \pm 1.3) \mbox{~pb}^{-1}$. HERA was operated with
electrons or positrons of $27.5\mbox{~GeV}$. The proton beam energy was 
$820\,\mbox{~GeV}$ before 1998 and $920\,\mbox{~GeV}$ since.

\subsection{Detector and Event Selection}\label{sec:sel}
The experimental methods are similar to those described 
in~\cite{Aid:1996dn,Adloff:1999zs,susanne}. Details of the detector and the 
analysis  can be found in \cite{h1det} and \cite{Kruger:2001jr}, respectively.
$J/\psi$ mesons are detected via the decays $J/\psi \rightarrow \mu^+\mu^-$
(branching fraction of $(5.88\pm 0.10)\%$\cite{pdg.00}). 
Reconstructed tracks in the central tracking detector (CTD) 
are identified as muons either because they produce minimal ionisation 
in the main (liquid argon, LAr) calorimeter or because they are linked to a track 
in the instrumented iron return yoke of the magnet (muon detector).
The trigger requirements rely on signals from track chambers and the muon 
detector, so for consistency at 
least one muon has to be reconstructed in the muon detector.

The event sample is subdivided into three (partially overlapping) datasets
to facilitate specific analyses. The datasets differ in the kinematic
requirements for $z$, \wgp\ and the minimum muon momentum $p_{\mu}^{min}$ 
imposed in the selection (see
table \ref{tab:select}). Dataset~I is a medium $z$ selection, $0.3<z<0.9$.
Dataset~II emphasizes the low $z$ region and the selection imposes a restriction 
in the polar angle of the muons to suppress background at $z\lsim0.2$.
The $z$ range, $0.05<z<0.45$, is chosen to overlap with dataset~I.
The \wgp\ ranges for the datasets I and II are adjusted to give high 
acceptance in the corresponding $z$ range. This results in a reduced 
\wgp\ range for dataset~II. %: $120<\wgp<260\,\gev$.
In order to determine the differential cross section \dsdz\ 
in the full $z$ range, a third sample, dataset~III, is defined, which 
covers the same \wgp\ region as dataset~II but covers a complementary $z$ range identical to that of
dataset~I.  

\begin{table}[b!]\centering
\begin{tabular}{|l|r|r||r|r|}
\hline
Dataset & $z$  & $W_{\gamma p}$[\gev] & $\theta_\mu [^\circ]$& $p_{\mu}^{min}\,[\gev]$
\\ \hline
I&   0.3 - 0.9 & 60 - 240 & 20 - 160 & 1.1 \\
II&  0.05 - 0.45 & 120 - 260 & 20 - 140 & 0.8 \\
III&  0.3 - 0.9 & 120 - 260 & 20 - 160 & 1.1  \\
\hline
\end{tabular}
\caption{\it Overview of kinematic ranges and selection criteria for the 
three datasets. All datasets are restricted to $Q^2 < 1 {\rm GeV}^2$ and
$\ptpsi > 1 {\rm GeV}$.}
  \label{tab:select}
\end{table}

In selecting \jpsi\ events two identified muons with momenta 
$p_\mu>1.1\,\gev$  are required (datasets I and III) and 
$p_\mu>0.8\,\gev$ for dataset~II. One of the muons must be  
reconstructed in the muon detector and has to have a momentum 
$p_\mu>1.8\,\gev$. In addition to the decay muons at least three additional 
tracks are required in order to suppress diffractive
\jpsi\ contributions. Furthermore a cut on the transverse momentum of the muon 
pair, \ptpsi$>1\,\gev$, is applied.
%`good' track ($\ptr>150\,\mev$ and minimum track length of 12$\,$cm). 
%The  position of the reconstructed event vertex is required to be within the 
%nominal interaction region. 
The restriction to photoproduction is made by removing events with a cluster 
of energy above 8~GeV in the electromagnetic part of the calorimeter.
The mean \qsq\ is then $0.05\,\gevt$.

%-------------------------------------------------------------------------------
\begin{figure}
\unitlength1.0cm
\begin{picture}(16,7)
\put(2.5,5.65){a)}
\put(9.8,5.65){b)}
%\put(12.5,5.8){\bf\large H1}
\put(0,-1.){\epsfig{file=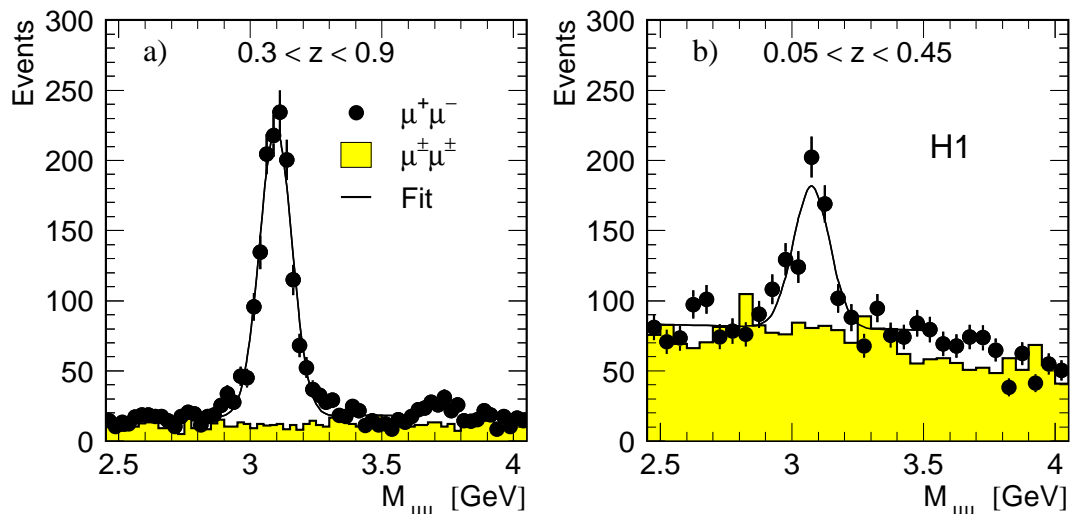,width=16cm}}
\end{picture}
\caption{\it Distribution of the invariant mass of the muon pairs
  after all selection cuts a) in the range $0.3<z<0.9$,
   $60<W_{\gamma p}<240\,{\rm GeV}$ and $p_{t,\psi}>1\,{\rm GeV}$
   (dataset~I) and b) in the range $0.05<z<0.45$, $120<W_{\gamma p}<260\,{\rm GeV}$ 
   and $p_{t,\psi}>1\,{\rm GeV}$ (dataset~II).  
   The data points with statistical error bars show the unlike sign 
   combinations, the 
   histograms the like sign combinations. The lines  are fits as
   described in the text.} 
\label{fig_signals}
\end{figure}

The distribution of the invariant mass of the two muon candidates after these 
cuts is shown for datasets I and II in Fig.~\ref{fig_signals}. 
There is a considerable non-resonant background in regions of low $z$ 
 due mainly to misidentified leptons as can be 
seen from the distribution of like sign pairs in 
Fig.~\ref{fig_signals}b. The following 
procedure is adopted to extract the number of 
signal events in a given analysis bin:  the mass distribution is fitted in the 
interval $2.45<M_{\mu\mu}<3.45\,\gev$ with a 
superposition of a Gaussian to describe the signal and a  function linear in mass to 
describe the background. The width and position of 
this Gaussian are determined from an overall fit to the data and are
then fixed. The number of signal events is obtained by 
counting the number of muon pairs in the interval $2.9<M_{\mu\mu}<3.3\,\gev$ 
and subtracting the fitted background in this interval. 
By varying the shape assumed for the background the
systematic error in the number of signal events is estimated to be 
5-15\% at low $z$ values and less than 1\% at higher $z$ values.  

\subsection{Kinematics} 
The following kinematic variables are used to describe 
\jpsi\ production: the square of the $ep$ centre of mass energy, $s= (k+p)^2$,
the negative squared four-momentum transfer $Q^2 = -q^2$ 
and the photon proton \cms\ energy \wgp$= \sqrt{(p+q)^2}$. Here $k$, $p$ and
$q$ are the four-momenta of the incident lepton, proton and
the exchanged photon, respectively. The scaled energy transfer 
$y=q\cdot p / k\cdot p $ is the energy fraction transferred from the lepton to the
hadronic final state in the proton rest frame. Neglecting the lepton and proton 
masses, a good approximation is $\wgpt \approx ys$ in the limit of $\qsq\simeq0$ considered here. 
The fraction of the  photon energy transferred to \jpsi\ meson 
in the proton rest frame is given by
$$z = \frac{p_\psi\cdot p}{q\cdot p},$$ 
where $p_\psi$ denotes the $J/\psi$ four-momentum.

The kinematic variables are reconstructed using the Jacquet-Blondel 
method~\cite{j-b}: 
\begin{eqnarray*}
 y&=&\frac{1}{2E_{e}} \sum_{had} \left(E-p_{z}\right),\\
z &=& \frac{(E-p_z)_{\jpsiw}}{ 2\,y\,E_e},
\end{eqnarray*} 
where $E_e$ denotes the energy of the incoming lepton. $(E-p_{z})_{\jpsiw}$ is 
calculated from the \jpsi\ decay muons and  $\sum_{had}$ 
runs over all final state hadrons including the \jpsi\ meson.
 In the calculation of the final state hadronic energy a
combination of tracks reconstructed in the CTD and energy depositions  
in the LAr and SpaCal calorimeters is used. 

The resolution for \wgp\ and \ptt\ is  much smaller than the width of the chosen 
analysis 
intervals. The absolute resolution for $z$ is typically $0.08$ for $z>0.4$ 
and $0.02$ at $z\sim0.1$. The size of the $z$ bins is chosen so 
that in the Monte Carlo simulation the fraction of events
reconstructed in the bin in which they
 were generated is above 60\%.

\subsection{Comparison of Data and Simulation}
The Monte Carlo program 
EPJPSI \cite{epjpsi} is used to generate two types of simulated datasets.
The first contains \jpsi\ events from direct photon gluon fusion,
the second from gluon gluon fusion corresponding to  
the  main expected contribution to the resolved photon component. Both
processes are calculated in the \csm\ in 
LO. Relativistic corrections are included, which enhance the cross section at 
high $z$ values~\cite{epjpsi}. 
The proton and photon gluon densities in the simulation are taken from 
MRS(A$^\prime$)\cite{MRS_Ap} and GRV-LO~\cite{GRV}, respectively.
The QCD renormalisation and factorisation scales
are chosen as \mpsit. Parton showers are 
simulated but their effect is small.

The procedure to extract cross sections relies on the detailed simulation of 
the H1 detector response. Particular care is taken to adjust the efficiency 
functions for the muon identification and trigger algorithms in order to 
achieve a reliable description as a function of the muon momenta and angles. This 
adjustment is performed using high statistics datasets from diffractive 
\jpsi\ production \cite{Kruger:2001jr}. Remaining differences between data and 
simulation are taken into account 
in the systematic error. 

\begin{figure}[p]
\unitlength1.0cm
\begin{picture}(16,8)
\put(1.5,9.5){a)}
\put(6.9,9.5){b)}
\put(14.7,9.5){c)}
\put(-0.75,4.7){\epsfig{file=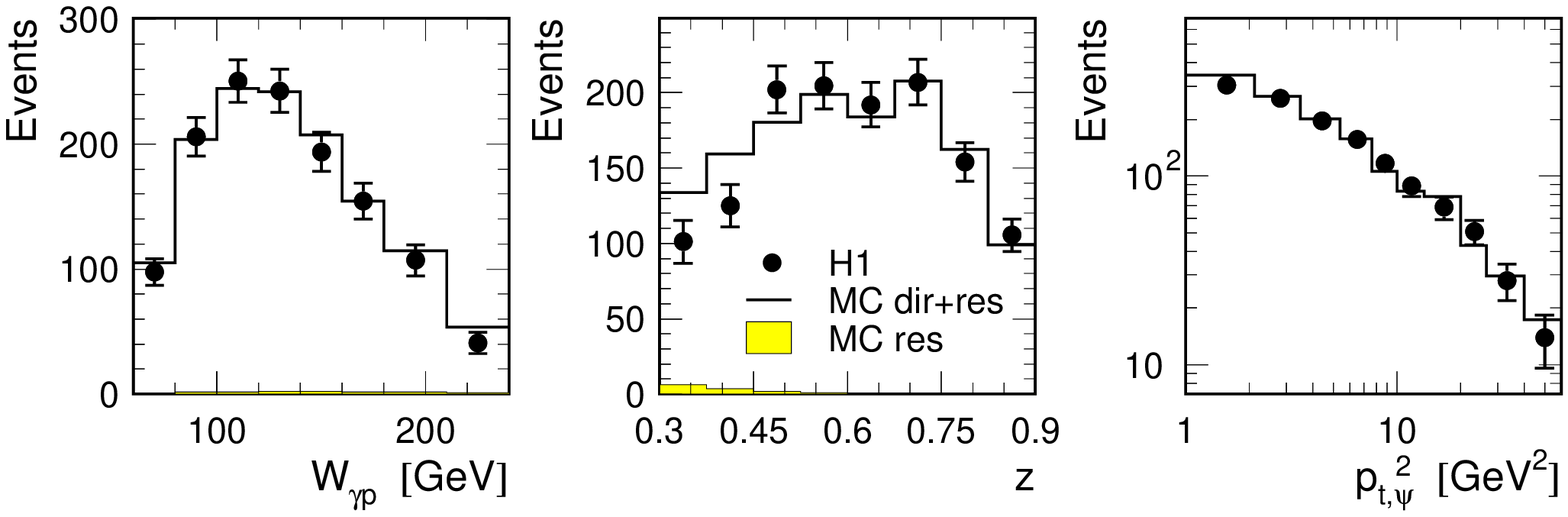,width=17.5cm}}
\put(-0.75,-0.7){\epsfig{file=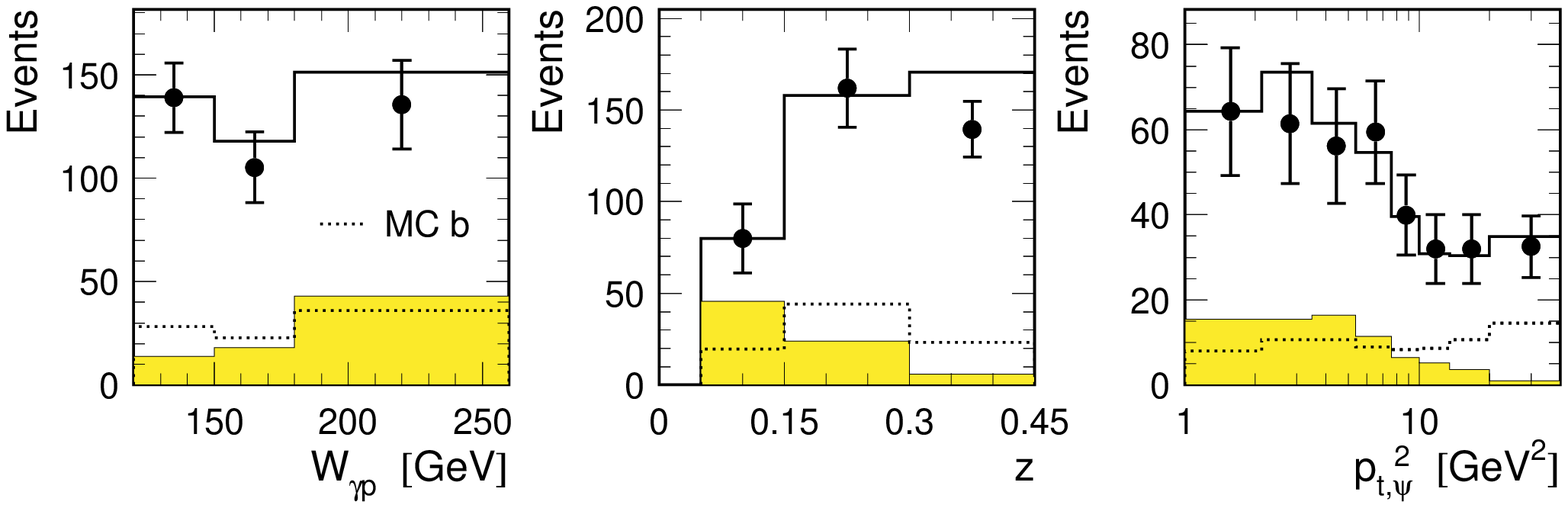,width=17.5cm}}
\put(1.9,4.1){d)}
\put(7.,4.1){e)}
\put(14.7,4.1){f)}
\end{picture}
\caption{\it Comparison between data and Monte Carlo 
simulations after all selection cuts and after subtraction of non-resonant
background. 
a) \wgp, b)  $z$ and c) $p_{t,\psi}^2$ distributions in the range $0.3<z<0.9$, 
$60<W_{\gamma p}<240\,{\rm GeV}$ and $p_{t,\psi}^2>1\,{\rm GeV}^2$ (dataset~I).
d)  \wgp, e)  $z$ and f) $p_{t,\psi}^2$  distributions in the range 
$0.05<z<0.45$, $120<W_{\gamma p}<260\,{\rm GeV}$ and
$p_{t,\psi}^2>1\,{\rm GeV}^2$ (dataset~II). 
  The error bars on the data points are statistical only. The EPJPSI simulation 
of direct + resolved photon processes (full histograms) and resolved photon
processes (shaded histograms) are compared with the data (for the
relative normalisation see text). An estimate of the direct photon 
contribution from  $b\ra\jpsiw+X$ from EPJPSI (which is not included in the full histogram) 
is also shown as a dotted histogram in d)--f).}
\label{fig_control_medz}
\label{fig_control_lowz}
\end{figure}

The two simulated datasets for direct and resolved photon 
processes are merged and normalised to the data, as follows 
(see also Fig.~\ref{fig_control_medz}a--c and d--f for the medium and low $z$ 
datasets I and II, respectively).
First, the generated events of the direct photon simulation are 
reweighted in \ptt\ according to a parametrisation of the measured 
distribution in the medium $z$ range. The simulation  
is then normalised to the data in the range $0.45<z<0.9$, where backgrounds  
from diffractive and resolved photon processes are expected to be negligible.
 Extrapolation of the simulation into the low $z$ region shows that 35 \jpsi\ 
mesons are expected in the interval $0.05<z<0.15$ whereas 80 \jpsi\ meson candidates 
are observed. 
%  (80\pm19) 
%In this way the
%total fraction of resolved photon processes is found to be $\sim20\%$ in dataset~II. 
This excess of observed events over the direct photon simulation is
attributed to the resolved photon component, which is normalised accordingly.
In principle a contribution from $\bbbar$\ pairs decaying to
\jpsi\ mesons should also be taken into account. However, the theoretical 
and experimental uncertainties are still large for $\bbbar$\ production, 
and the process is thus neglected here (see also discussion in the next section).

%Adding the simulated resolved photon contribution to compensate for the difference 
A contribution of resolved photons of about
$50\%$ at $0.05<z<0.15$ is obtained falling to 10\% for
$0.15<z<0.3$ and being negligible for $z>0.3$ (Fig.~\ref{fig_control_lowz}e).
Note that the resolved component is not reweighted in \ptt\
since good agreement of the superposition 
of the two simulations with the data is found (Fig.~\ref{fig_control_lowz}f).
This may, however, be fortuitous because of the neglected \bcon, which 
is expected to be important particularly at high values of
\ptt. Current simulations (AROMA \cite{aroma} and EPJPSI \cite{epjpsi}) normalised 
to the measured value for $ep \ra e b{\bar b} X$ with \qsq$<1\,\gevt$ of 
$(14.8 \pm 1.3 ^{+3.3} _{-2.8})\mbox{nb}$~\cite{H1_b}, predict $b$ contributions 
of the order of $50\%$ or more in the highest analysis interval 
$20<$\ptt$<40\,\gevt$.

\subsection{Other Contributions to \boldmath\jpsi\ Production}\label{sec:back}

After the  selection cuts  described in section \ref{sec:sel} the 
\jpsi\ sample is dominated by inelastic \jpsi\ production, 
in which the \jpsi\ is directly produced from the \ccbar\ pair 
in the boson gluon fusion process. 
However, contributions remain from both the diffractive and inelastic 
production of \psits\ mesons, the production of $b$ flavoured
hadrons and possibly also from $\chi_c$ states.
The estimated amounts of these feeddown processes are summarised in Table~\ref{tab:background}. 
Note that these contributions are {\em not} subtracted, either because they 
are negligible, such as diffractive backgrounds, or, because  
the dependence on the kinematic variables has not been measured
and is only poorly known from theory, as in production of $\bbbar$\ pairs 
with subsequent decays to \jpsi\ mesons.  
  
\begin{table}[b!]
\begin{center}
\begin{tabular}{|l|c|c|} 
\hline
%\multicolumn{3}{c}{}\\ \hline
%\multicolumn{3}{|c|}{Resonant background}\\
\multicolumn{1}{|c}{Source}& \multicolumn{2}{|c|}{Contribution (\%)}\\ 
 & Dataset~I, III  & Dataset~II \\ \hline \hline
Diffractive $\psi(2S)\ra \jpsiw+X$ ($0.75<z<0.9$) & $1\,$ ($3$) & - \\ \hline
Inelastic $\psi(2S)\ra \jpsiw+X$ & $15$ & $15$ \\ \hline 
$b\ra\jpsiw+X$ ($0.3<z<0.45$) & $5\,$ ($19$) & $25$\\ \hline
$\chi_c \ra \jpsiw+\gamma$ & $1$ & $7$ \\ \hline
\end{tabular}
\caption{\it Estimated contributions from other processes to the measured 
\jpsi$+X$ data sample in addition to inelastic boson 
gluon fusion. The amounts in parentheses refer to the $z$ bins in parentheses. }
\label{tab:background}
\end{center}
\end{table}

Contributions from diffractively produced $J/\psi$ and $\psi(2S)$ mesons are
effectively suppressed by the requirements of at least three tracks in 
addition to the decay muons and $\ptt>1\,\gevt$. 
The remaining contribution from the cascade
decay of $\psi(2S)$ is estimated with a simulation of diffractive 
$\psi(2S)$ production using the DIFFVM generator program~\cite{diffvm}. 
DIFFVM is normalised to the observed signal of $\psi(2S)\ra\jpsiw\pi^+\pi^-$ 
in a sample of events with exactly four tracks.
A residual contribution from diffractive $\psi(2S)$ production 
of about $3\%$ in the highest $z$ bin is inferred. Averaged over the 
total $z$ range this background is less than $1\%$ in the medium $z$ region and 
negligible at low $z$.

The inelastic photoproduction of $\psi(2S)$ mesons with subsequent
decays to \jpsi\ mesons is expected to contribute 
$15\%$ to $J/\psi$ production~\cite{csm_nlo} and to show similar dependences 
on the kinematic variables because the production processes are the same.

The contribution from the decay of $b$ flavoured hadrons can only be 
estimated with large uncertainties, since few measurements exist \cite{H1_b} and 
these considerably exceed current theoretical estimates. Using the
EPJPSI generator program and normalizing the production cross section to 
the measured value for $ep \ra e b{\bar b} X$ with \qsq$<1\,\gevt$~\cite{H1_b}, 
a contribution of
$(25 \pm6)\%$ due to \bdec\ is estimated in the kinematic region of
%$25 ^{+6} _{-5}\%$
dataset~II (see also Fig.~\ref{fig_control_lowz}). At higher $z$ and lower 
\wgp\ the background fraction is smaller. 
For example in the lowest $z$ bin of dataset~I a contribution of $(19\pm5)\%$ 
%$19 ^{+5} _{-4}\%$ 
is found which would correspond to $5 \pm 1\%$ in the total $z$ region of dataset~I.  

A further potential contribution comes from  $\chi_c$ decays,  
$\chi_c\ra\gamma+\jpsiw$. %yielding $J/\psi$ distributions very similar
%to those where \jpsi\ mesons are produced in the hard interaction process. 
The \csm\ predicts that $\chi_c$ mesons can only be produced in resolved 
photon processes due to quantum number conservation. 
From EPJPSI $7\%$ of the total resolved contribution is 
estimated to arise from this source. The production of $\chi_c$ mesons
is possible also in direct photon gluon fusion via \coloct\ $^3S_1$ states,
%The cross section ratio depends on the ratio of the LDMEs
%${\cal O}^{\chi_{c0}}$[$8,^3S_1] / {\cal O}^{J/\psi}$[$1,^3S_1]$.
%With a choice of ${\cal O}^{\chi_{c0}}$[$8,^3S_1] \sim 0.3 \times 
%10^{-2}\,{\rm GeV}^3$
which would result in a contribution of less than $1\%$ in the medium $z$ region~\cite{kraemer}.

%((KATJA) MUSS man hier den z Bereich angeben? A: Nein, beide Prozesse haben das
%gleiche Feynman-Diagramm, unterschiedlich sind nur das zusaetzliche Photon vom
%Zerfall und die weichen Gluonen. )

\subsection{Systematic Uncertainties}
The estimated systematic errors on the cross sections are listed in 
Table~\ref{tab_syserror}.
The uncertainty from the track reconstruction is based on an error of 2\%
for the efficiency
per track and takes into account the observed track multiplicity distribution. 
The systematic error of the total hadronic energy includes 
the uncertainty of the calorimeter energy scales (4\% for the LAr, 
and 10\% for the SpaCal) and leads to cross section error estimates 
of 4\% (6\%) in the medium (low) $z$ regions.
The systematic errors in the muon identification and trigger efficiencies 
are estimated from the residual differences between data and simulation.
The error in the determination of the event numbers in the analysis
bins is estimated by changing the description of the non-resonant
background from a linear dependence on the invariant mass to a power law.
The error on the acceptance due to restricting 
the polar angle of the decay muons is obtained by 
using different proton and photon parton densities which influence
the \wgp\ distribution. In general the changes in the acceptance are small. 
A significant uncertainty is found only at large \wgp\ in
dataset~I and at small \wgp\ in dataset~II.
The uncertainty in the acceptance at low $z$ due to the modelling of the $z$ 
dependence is found to be 4\%.
The low $z$ dataset~II contains a sizeable contribution from \bdec.
The effect on the acceptance is estimated using the EPJPSI simulation
of $\bbbar$\ production with a relative contributions as in table \ref{tab:background}.
%In addition to the expected shift by 
%$\sim 2.5\%$ percent changes $4\%$ are observed. 
The decay angular distribution of the \jpsi\ meson has an influence on
the acceptance mainly in the medium $z$ region. 
Varying the decay angular distribution of the \jpsi\
meson within the measured uncertainties\footnote{The parameter $\lambda$ in
equation~(\ref{eq:th}) in section \ref{sec:pol} is varied from 0 (used in the 
simulation) to $+1$ or $-0.5$.} changes the acceptance by $\pm8\%$ nearly 
independently of other variables.
\begin{table}[h!]
\begin{center}
\begin{tabular}{|l|c|c|} 
\hline
%\multicolumn{3}{|c|}{Systematic Errors}\\
\multicolumn{1}{|c}{Source} & \multicolumn{2}{|c|}{Uncertainty [\%]}\\ 
 & Dataset~I, III & Dataset~II \\ \hline \hline
Track reconstruction efficiency& $6$ & $6$\\ \hline
Muon identification efficiency& $3$ & $7$ \\ \hline
Calorimeter energy scales & $4$ & $6$ \\ \hline
Signal events & $<1$ & $5-15$\\ \hline
%L1 iron trigger efficiency & $2.6$ & $1.9$\\ \hline
%L1 track trigger efficiency & $3.8$ & $9.7$\\ \hline
%L2NN efficiency & $2.8$ & - \\ \hline
%L4 trigger efficiency & $3$& $3$  \\ \hline
Trigger efficiency & $6$ & $10$\\ \hline
%parton density functions ($210<W<240\gev$) & $4$ ($+8$) & $5$ ($10$) \\ \hline \hline
Acceptance: PDFs  & $4-8$ & $5-10$ \\ \hline
\hspace{2.1cm} $z$ dependence & - & $4$\\ \hline
\hspace{2.1cm} \jpsi\ decay ang. distr.& $8$ & $7$ \\ \hline 
\hspace{2.1cm} $b\ra\jpsiw+X$& $<1$ & $3-12$ \\ 
\hline \hline
Integrated luminosity & \multicolumn{2}{|c|}{$1.5$}\\ \hline
Branching ratio & \multicolumn{2}{|c|}{$1.7$}\\ \hline\hline %5.88+-0.10
Total & $13.3-15.5$ & $18.9-24.3$ \\ \hline
\end{tabular}
% medium z $210<W<240$: +12.4 -9.5
% medium z $0.75<z<0.9$: +10.2 -10.7
% low z $180<W<260$: $17.6$
\caption{\it Systematic uncertainties for the production cross
sections in the medium $z$ and low $z$ regions. The error due to the 
\jpsi\ decay angular distribution does not apply to the polarisation 
analysis. }
\label{tab_syserror}
\end{center}
\end{table}

\section{Results}

Cross sections for $ep\ra e\jpsiw X$
are converted to $\gamma p$ cross sections using the 
equivalent photon approximation~\cite{epa,fri}. 
The photon proton cross section is defined by: 
$$%\begin{equation}
d\sigma_{ep}=\sigma_{\gamma p}\:f_{\gamma/e}(y) dy\mbox{,}
%\label{formel_ep_gp}
$$%\end{equation}
where $f_{\gamma/e}$ denotes 
the photon flux\footnote{$f_{\gamma/e}(y) = \frac{\alpha}{2 \pi} 
\left( 2m_e^2 y \left(\frac{1}{Q^2_{min}}-\frac{1}{Q^2_{max}}\right)
+\frac{1+ (1-y)^2}{y}\log{\frac{Q^2_{max}}{Q^2_{min}}}\right)$, where $Q^2_{min}=m_e^2\frac{y^2}{1-y}$, taken from \cite{fri}.}
integrated over \qsq\ from the kinematic limit of $Q^2_{min}$
to the upper limit of the measurement, $Q^2_{max}=1\,\gevt$.
The measured cross sections are given at the weighted mean 
for each bin assuming specific functional forms to fit the data, i.e. 
\mbox{$\sigma_{\gamma p}(\wgp)\propto\wgp^\delta$}  and 
\dsdptt$\propto\mbox{$(\ptt+M_\psi^2)^{-n}$}$ as described below.

The measured cross sections are displayed in Figs.~\ref{fig1}--\ref{cascade} 
and the numerical values are collected in 
Tables~\ref{tab_medz_xsec_pt}--\ref{tab_allz_xsec}. 
In the next sections the measured cross sections are compared with
theoretical calculations (Table~\ref{tab:model1}). 

\subsection{Comparison of Cross sections with the Colour Singlet Model}
\begin{figure}[htbp]
\unitlength1.0cm
\begin{picture}(16,18)
\put(2.5,12.7) {a)}
\put(10.,12.7) {b)}
\put(2.5,5.7) {c)}
\put(0,-1){\epsfig{file=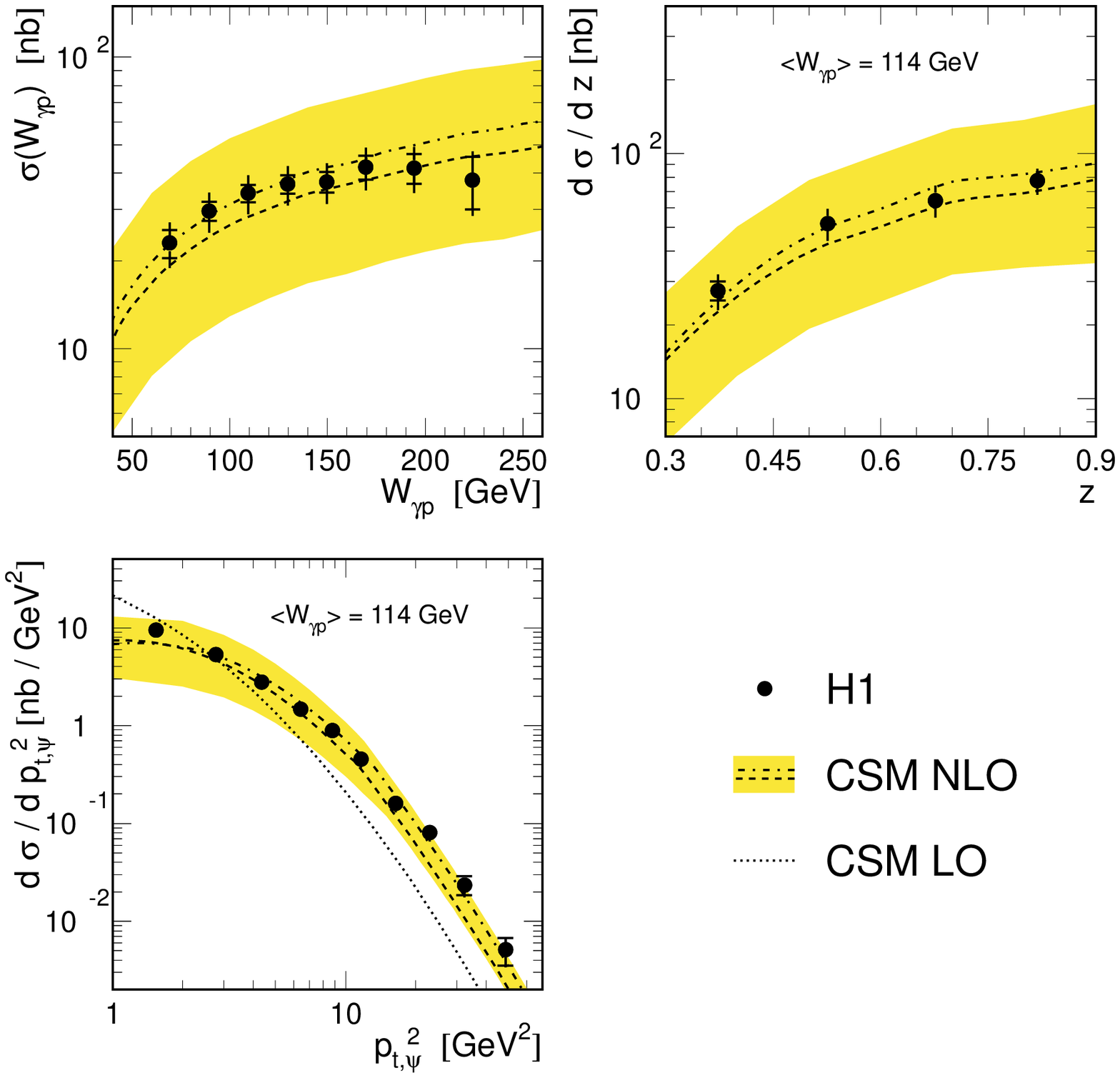,width=16cm}}
\end{picture}
\caption{\it Inelastic \jpsi\ production in the region $60<W_{\gamma p}<240\,\gev$, 
$0.3<z<0.9$ and \ptt$>1\,\gevt$. a) Total cross section as a function of \wgp,
b) differential cross sections ${\rm d}\sigma/{\rm d}z$  and 
c) ${\rm d}\sigma/{\rm d}p_{t,\psi}^{2}$. 
The H1 data are shown together with NLO calculations in
the \csm~\cite{csm_nlo} with MRST~\cite{mrst} parton density functions 
(\wgp$=115\,\gev$ for b) and c)). The shaded band reflects the
uncertainties in $m_c$ and $\alpha_s$ (see Table~\ref{tab:qcd}); 
the dashed (dash-dotted) curve %close to the data 
is calculated with  $m_c = 1.3 (1.4)\,{\rm GeV}$,
$\alpha_s(M_Z) = 0.1175 (0.1225)$. 
In c) a CSM LO calculation with MRST parton distributions, $m_c = 1.3\,{\rm GeV}$, 
$\alpha_s(M_Z) = 0.1225$ is also shown (dotted).}
 \label{fig1}
\end{figure}

\begin{figure}
\unitlength1.0cm
\begin{picture}(16,11)
%\put(7,13.){\Large Medium $z$}
\put(1,-0.2){\epsfig{file=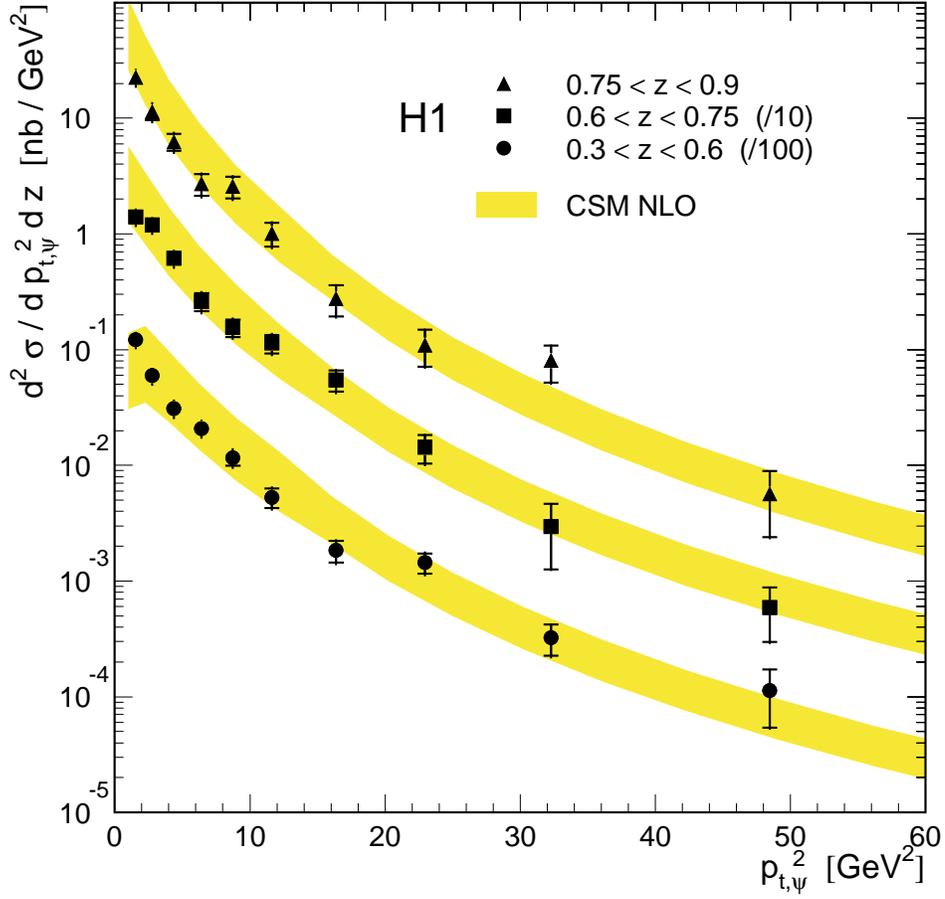,width=14cm}}
\end{picture}
\vspace{-1.2cm}
\caption{\it Double differential cross section ${\rm d}^2\sigma/
{\rm d}p^{2}_{t,\psi}\,{\rm d}z$ in the same kinematic region
and 
%$60<W_{\gamma p}<240\,{\rm GeV}$, 
%$0.3<z<0.9$. The shaded band shows the result of 
compared with the same CSM NLO calculations as in Fig.~\ref{fig1}.}
%The dashed lines are fits of the form $a \cdot (p^{2}_{t,\psi}+M_\psi^2)^{-n}$.}
\label{double}
\end{figure}

The photon proton cross sections \sgp, 
\dsdz\ and \dsdptt\ in the medium $z$ range are given in Figs.~\ref{fig1} and \ref{double} and Table~\ref{tab_medz_xsec_pt}. 
The data are first compared with the NLO CSM calculations \cite{kraemer}. 
Although the \ptt\ range of the measurement has been considerably
enlarged compared with earlier analyses the agreement between data and theory remains  
good (Fig.~\ref{fig1}c). 
The band on the theoretical calculations in 
Figs. \ref{fig1} and \ref{double} 
represents the uncertainties due to the variations of the mass of the charm quark, 
$m_c =1.4\pm0.1\,{\rm GeV}$ and of 
$\alpha_s(M_Z)=0.1200\pm0.0025$ \cite{pdg.00} (see Table~\ref{tab:qcd} 
for the other parameter values). 
In the \wgp\ and $z$ distributions of Fig. \ref{fig1}a and b these
uncertainties affect mainly the normalisation. 
The shapes and normalisations of the calculated cross sections as functions of 
\wgp\ and $z$ are in approximate agreement with the data for the two  
parameter sets, $(m_c,\,\alpha_s) = (1.3\,\gev,\,0.1175)$ and  
$(1.4\,{\rm GeV},\,0.1225)$. 
The combinations $m_c=1.3\,\gev$ and $\alpha_s(M_Z) = 0.1225$ 
(upper limit of the band) 
and $m_c=1.5\,\gev$ and $\alpha_s(M_Z) = 0.1175$ (lower limit) can be excluded
here.

The NLO CSM calculation describes  the \ptt\ distribution rather well 
(Fig.~\ref{fig1}c).
This is not the case for the LO CSM calculation, 
%(MRST parton densities, $m_c = 1.3\,{\rm GeV}$ and $\alpha_s(M_Z) = 0.1225$)
 which lies above the data at low \ptt\ and 
falls too steeply towards higher values of  \ptt. 
The \ptt\ dependence is sensitive to the choice of $m_c$ and $\alpha_s$.
The calculations with $(m_c,\,\alpha_s) = (1.3\,\gev,\,0.1175)$ and 
$(1.4\,{\rm GeV},\,0.1225)$, with which the \wgp\ and $z$ distributions are well 
described, also give a reasonable description 
of the \ptt\ dependence apart from a tendency to undershoot the data at high \ptt. 
The \ptt\ distribution in bins of $z$ (Fig.~\ref{double} and 
Table~\ref{tab_medz_xsec_double}) is also described well by 
the NLO CSM calculation. This is interesting because the size of the  
\coloct\ contributions in the NRQCD approach is predicted to depend on $z$.

\subsection{Parametrisations of Transverse Momentum Distributions} 
The differential cross sections $d\sigma/d\ptt dz$ (Fig.~\ref{double} and 
Table~\ref{tab_medz_xsec_double}) in the different $z$ regions 
are found to be well described by a functional form $(\ptt+M_\psi^2)^{-n}$. 
Fits to the data have been performed and the results for the exponents $n$ are 
summarised in Table~\ref{tab_n}.   %(total errors)
There is a tendency for $n$ to decrease towards lower $z$ values.
A similar analysis has been carried out for inelastic \jpsi\
production in the range $2<\qsq<100\,\gevt$  ($0.3<z<0.9$, 
$50<\wgp<225\,\mbox{~GeV}$)~\cite{susanne}. There the differential cross
sections as a function of the transverse momentum in the photon proton \cms\ 
system were fitted to the same functional form yielding a value 
$n=4.15\pm0.50$, which agrees very well with the present result.

\begin{table}[hb]
\begin{center}
\begin{tabular}{|l|c|c|} 
\hline
\wgp[$\gev$]&$z$&$n$\\ \hline
$60-240$& $0.3-0.9$ &$4.49\pm0.15$\\\hline
$60-240$      & $0.75-0.90$ &$4.8\pm0.2$\\ 
      & $0.6-0.75$  &$4.6\pm0.2$\\ 
      & $0.3-0.6$   &  $4.4 \pm 0.2$\\\hline 
$120-260$ & $0.05-0.45$ & $4.1\pm0.2$\\
\hline
\end{tabular}
\end{center}
\caption{\it Results of fits to differential cross sections $d\sigma/d\ptt$ of the form 
$(\ptt+M_\psi^2)^{-n}$ in different kinematic regions. Total experimental 
errors (statistical and systematic added in quadrature) have been used in the fits.} 
\label{tab_n}
\end{table}

\subsection{Comparison of Cross sections with NRQCD Calculations}

\begin{figure}[htbp]
\unitlength1.0cm
\begin{picture}(16,7.7)
\put(2.5,6.5) {a)}
\put(10.,6.7) {b)}
\put(-1.,-0.7){\epsfig{file=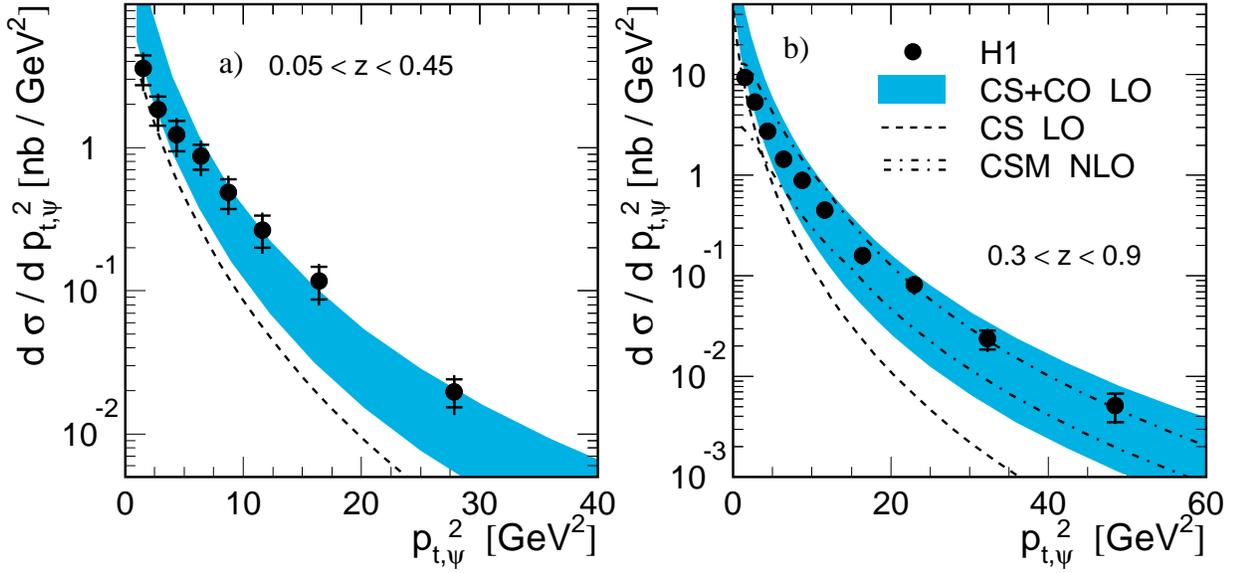,width=18cm}}
\end{picture}
\caption{\it Differential cross section ${\rm d}\sigma/{\rm d}p_{t,\psi}^{2}$
for inelastic \jpsi\ production in two kinematic regions. 
a) $0.05<z<0.45$, $120<W_{\gamma p}<260\,{\rm GeV}$; 
b) $0.3<z<0.9$, $60<W_{\gamma p}<240\,{\rm GeV}$ (same data as in
Fig.~\ref{fig1}c). 
The data are compared with LO theoretical calculations in the NRQCD 
framework, including \coloct\ and \colsing\ contributions from direct and resolved 
photons shown as the shaded band, which reflects the uncertainties due to the LDMEs. 
The dashed line shows the \colsing\
contribution separately. In b) the band of the same CSM NLO calculation
\cite{csm_nlo} as in Fig.~\ref{fig1}c is also shown as dash-dotted lines.} 
\label{cslowz2}
\end{figure}

\begin{figure}[htbp]
%\unitlength1.0cm
%\begin{picture}(16,9.)
%\put(0,-1.0){
\epsfig{file=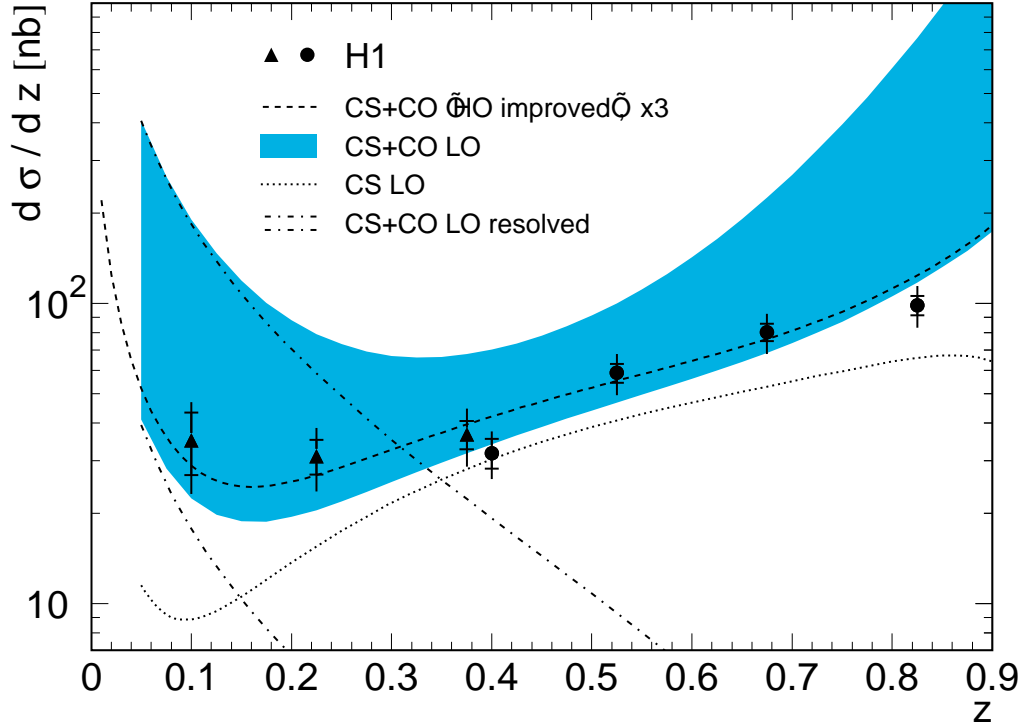,width=15cm}
%}
%\end{picture}
\caption{\it Differential cross section ${\rm d}\sigma/{\rm d}z$ 
for $120<W_{\gamma p}<260\,{\rm GeV}$ and $p^2_{t,\psi}>1\,{\rm GeV}$.
The data are shown as triangles (points) corresponding to datasets~II
(III). The two data points from different analyses at $z\approx0.4$ are statistically correlated and offset
for visibility. Theoretical 
calculations within the
NRQCD/factorisation framework including \coloct\ and \colsing\ contributions 
from direct and resolved photons are shown for comparison. The shaded band is 
a calculation by Kr\"amer \cite{kraemer} reflecting the uncertainties due to the LDMEs
(see Table~\ref{tab:qcd}). The two dash-dotted lines indicate the
region of the resolved contributions (CS+CO) separately and the dotted
line shows the \colsing\ contribution.
The dashed line is the result of a NRQCD calculation by Kniehl et
al. \cite{kniehl}, where higher order effects have been estimated and
which has been normalised to the data. Note that the theoretical calculations 
are for charm only, while the data contain a residual background from $b$ 
quarks at low~$z$.}
\label{allz}
\end{figure}

\begin{figure}[htbp]
\unitlength1.0cm
\begin{picture}(16,6.5)
\put(2.2,6.2) {a)}
\put(14.,6.2) {b)}
\put(0,-0.5){\epsfig{file=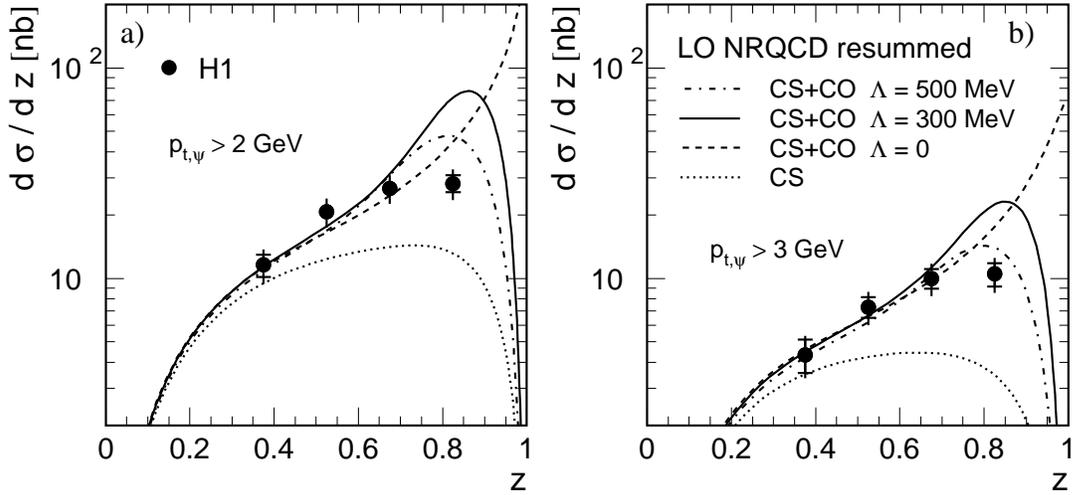,width=16cm}}
\end{picture}
\vspace{-1cm}
\caption{\it Differential cross sections ${\rm d}\sigma/{\rm d}z$ 
 ($60<W_{\gamma p}<240\,{\rm GeV}$) for a) $p_{t,\psi}>2\,{\rm GeV}$ and
b) $p_{t,\psi}>3\,{\rm GeV}$ in comparison with NRQCD calculations including \coloct\ 
and \colsing\ contributions
after resumming soft contributions at high $z$ \cite{wolf}. 
The curves correspond to three values of the parameter $\Lambda=0$ 
(no resummation), $300$ and $500\,{\rm MeV}$ (dashed, full and dash-dotted lines). 
The dotted line is the colour singlet 
contribution alone. The theoretical curves have been scaled with a
common factor 2 in a) and 3 in b), respectively.} 
\label{wolf}
\end{figure}

The comparison of the measured cross sections \dsdptt\ and \dsdz\  with the 
calculations within the NRQCD approach are shown in Figs.~\ref{cslowz2}--\ref{wolf}. 
For this comparison data in the low $z$ region are included 
(Tables~\ref{tab_lowz_xsec_pt} and \ref{tab_allz_xsec}).
There are considerable uncertainties in  such LO NRQCD calculations. 
The band in Figs.~\ref{cslowz2}  and \ref{allz} gives an estimate 
from \cite{kraemer} of the 
main uncertainty, which arises from the LDMEs obtained from
\jpsi\ production in \ppbar\ interactions. For the upper limit of the
 band the LDMEs extracted in LO are used, while for the lower limit  
values including estimates of higher orders in the extraction are used\cite{sanchis}. 
The distributions 
of \ptt\ are shown in Fig.~\ref{cslowz2}a and b for the low and medium $z$ 
ranges, respectively. In both $z$ regions the NRQCD calculation \cite{kraemer} and 
data are compatible within experimental and theoretical uncertainties. 
However there seems to be a difference in shape: 
the data have a slightly harder spectrum than predicted.
For the low $z$ range one 
should keep in mind that the calculation is for charm only, while the data 
contain feeddown from $b$ quarks, which is expected to contribute a harder 
transverse momentum distribution. 
In both $z$ regions the \colsing\ contribution alone falls significantly 
faster than the data. 

The differential cross section \dsdz\ extending over the full $z$ range, 
$0.05<z<0.9$, is compared with the same LO  NRQCD
calculation \cite{kraemer} in Fig.~\ref{allz} for a cut \ptt$>1\,\gevt$.
Resolved contributions are found to dominate in the calculation 
below $z\lsim0.15\, (0.3)$ depending on the choice of LDMEs. 
In the comparison of data and theory one should note that the
contribution from $b$ decays in the data is sizeable at low $z$ 
(of the order of 25\%). Nevertheless one may infer from Fig.~\ref{allz} 
that the LO NRQCD calculation, including CS and CO contributions, is able 
to give a fair description of the data both in shape and in normalisation, 
if the  LDMEs are chosen to be 
close to the lowest available estimates (lower edge of the shaded band). 
Using larger LDME values in this calculation leads to a strong increase of the
theoretical cross sections at high and low $z$ values which is inconsistent 
with the measurements. 
The \colsing\ contribution alone, which is also shown separately in Fig.~\ref{allz},
is roughly 30\% below the data for $z\gsim 0.5$ although the shape is
adequately described in this region. At lower $z$ values the \colsing\
contribution falls below the data by up to a factor 3.

Although the uncertainties in the calculations are substantial, 
the measured cross section in
the lowest $z$  bin, $0.05<z<0.15$, suggests that a resolved photon contribution is
present: Correcting for $b$ and inelastic $\psi(2S)$ feeddown using the estimates in 
Table \ref{tab:background} (section \ref{sec:back}) and neglecting their 
uncertainties the measured cross section is found to be 2 standard deviations above 
the direct photon contributions alone (including CS and CO contributions with a
range of LDME values as in Table~\ref{tab:qcd}\cite{kraemer}).

A different calculation in the NRQCD framework 
has been carried out by Kniehl et al.~\cite{kniehl} and is shown in
Fig.~\ref{allz} as 'HO improved'. 
In this calculation higher order effects were taken 
into account approximately  using NLO parton density 
distributions for the photon and proton and using the LDMEs corrected for 
estimated higher orders effects
(the parameters of the calculation are given in Table~\ref{tab:qcd}). 
This calculation gives a good description of the shape of the data but
a normalisation factor of 3, as suggested by the authors~\cite{kniehl}, 
is necessary to reconcile the predicted cross section with the data.

The tendency for the \coloct\ contributions to rise towards high $z$ values 
(upper end of shaded band in Fig.~\ref{allz}) is at variance 
with the data as has also been noted previously \cite{Aid:1996dn,zeus97_jpsi}. 
This discrepancy may be due to phase space limitations at the upper limit of $z$ 
where the emission of soft gluons in the conversion of the \ccbar\ pairs to
\jpsi\ mesons is suppressed. This is not taken into account in \cite{kraemer,kniehl}.
 In \cite{wolf} a resummation of the non-relativistic
expansion was carried out, leading to %`shape functions' which cause
a decrease of the predicted cross section at high $z$. 
In Fig.~\ref{wolf} (Table~\ref{tab_medz_xsec_pt}) the measured cross sections 
\dsdz\ for $\ptpsi>2\,\gev$ and for $\ptpsi>3\,\gev$ are compared with the results 
of these resummed calculations in the kinematic 
region of dataset I. 
The calculated curves have been roughly  normalised to the data points at low $z$. 
The resummed calculation for $\Lambda=500\,\mev$ gives an acceptable description of 
the data\footnote{The parameter $\Lambda$ describes the energy
lost by the \ccbar\ system, in its rest system, in the conversion to 
the \jpsi\ meson.}  at \ptpsi$ > 3\,\gev$ while the agreement between data and 
calculation is still poor for \ptpsi$ > 2\,\gev$ or for lower $\Lambda$ values.

%$M_{3.1} = 1.5 \cdot 10^{-2}GeV^{3}$
%$O^8(^3S_1) = 0.5-1.0 \cdot 10^{-2}GeV^{3}$
% Kraemer (review):
%$O^8(^1S_0) = 1.2 \cdot 10^{-2}GeV^{3}$
%$O^8(^3P_0)/m_c^2 = 1.2 \cdot 10^{-2}GeV^{3}$
%\rightarrow  M_{3.1}=4.9 \cdot 10^{-2}GeV^{3}$
%\paragraph(CASCADE}

\subsection{Comparison of Cross sections with \boldmath$k_t$ Factorisation Calculations}
\begin{figure}[htbp]
\unitlength1.0cm
\begin{picture}(16,14)
\put(6.5,11.2) {a)}
\put(9.5,11.2) {b)}
\put(6.5,5.) {c)}
\put(1,-0.75){\epsfig{file=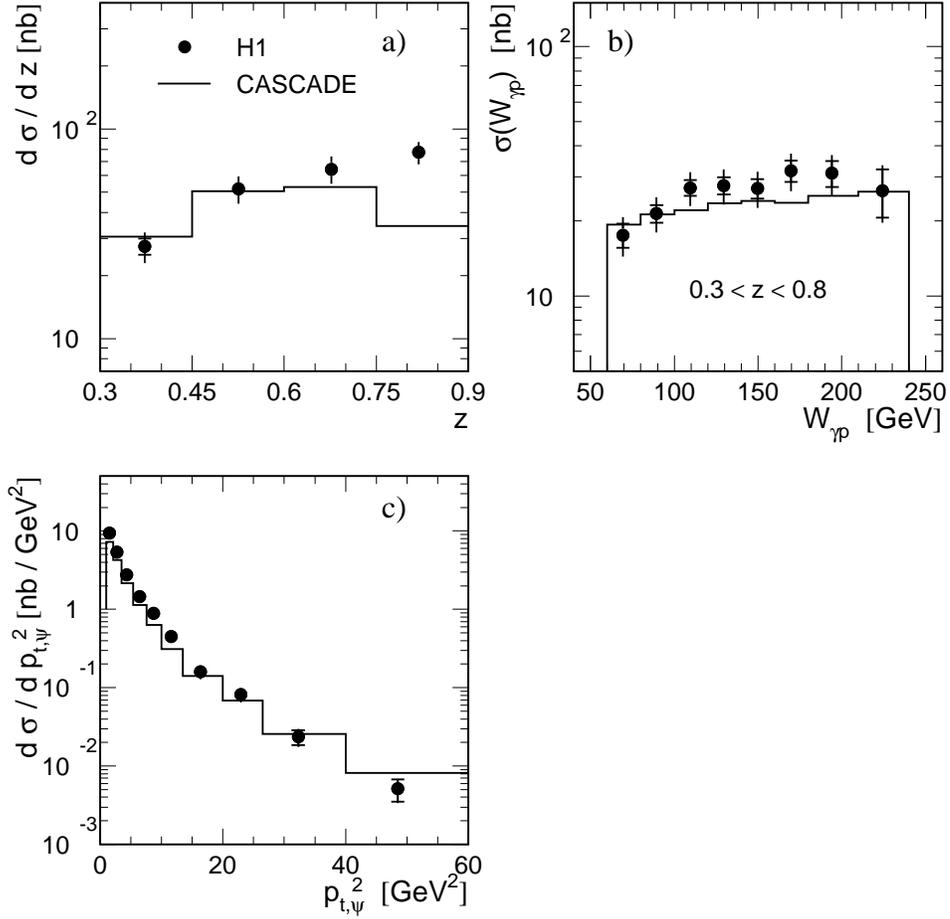,width=14cm}}
\end{picture}
\caption{\it Inelastic \jpsi\ production in the region $60<W_{\gamma p}<240\,\gev$, 
$0.3<z<0.9$ and \ptt$>1\,\gevt$ in comparison with a $k_t$
factorisation model implemented 
in the Monte Carlo generator CASCADE\cite{cascade}. 
a) Differential cross section ${\rm d}\sigma/{\rm d}z$,
b) $\sigma_{\gamma p}$ as a function of \wgp\ for $0.3<z<0.8$ and
c) \dsdptt.}
%\ and d) Ratio of data over simulation from CASCADE as a function of \ptt.}
\label{cascade}
\end{figure}

An entirely different approach to  inelastic \jpsi\ production within
the CSM\cite{Saleev:1994fg} 
is implemented in the Monte Carlo program CASCADE~\cite{cascade}.
Here direct photon gluon fusion processes are computed in the $k_t$ factorisation 
approach\cite{kt3,kt4} using an unintegrated 
($k_t$ dependent) gluon density in the proton and the gluons may thus be off-shell. 
This gluon density was 
obtained from a fit \cite{cascade} to the HERA structure function data 
(using the CCFM parton evolution equations).
Fig.~\ref{cascade} shows a comparison of the data with the results
from CASCADE.   In Fig.~\ref{cascade}a $d\sigma/dz$  is shown
as a function of $z$. Good agreement is observed between data and
predictions for $z\lsim 0.8$. At high $z$ values the CASCADE 
calculation underestimates the cross section. This may be due to missing 
relativistic corrections which are not available for the off-shell matrix
element. It could also indicate a possible missing CO contribution.
The \wgp\ distribution is shown for the restricted $z$ range $0.3<z<0.8$ 
(Fig.~\ref{cascade}b) and the CASCADE model is found to be in reasonable agreement 
with the data.
An important improvement compared to collinear LO calculations is
visible in the \ptt\ dependence (for $0.3<z<0.9$,
Fig.~\ref{cascade}c), where the CASCADE predictions fit the data quite well.
This is due to the transverse
momentum of the gluons from the proton which contributes 
to the transverse momentum of the \jpsi\ meson. 
%The ratio of data to
%simulation as a function of \ptt\ (Fig.~\ref{cascade}d) shows that the 
%CASCADE prediction is even slightly harder than the data. 

\subsection{Polarisation Measurement}
\label{sec:pol}
The polarisation of the \jpsi\ meson is expected to differ in the
various theoretical approaches discussed here and could in principle be used to 
distinguish between them, independently of normalisation uncertainties. 
The  polarisation analysis is performed in the ``target frame'', the rest 
system of the \jpsi\ meson, using the direction opposite to that of the
incoming proton 
%in the laboratory system 
as reference axis $z'$. Two angles are defined: $\theta^*$ is the polar 
angle of the positive decay muon with respect to the $z'$ axis and 
%$\mathit{\Phi}^*$  
$\Phi^*$ is the angle 
between the plane of the decay muons and the plane 
defined by the photon and the $z'$-axis.

\begin{figure}[htbp]
\unitlength1.0cm
\begin{picture}(16,4)
%\put(-1,-1){
%\epsfig{file=fig10.eps,width=18cm}
\epsfig{file=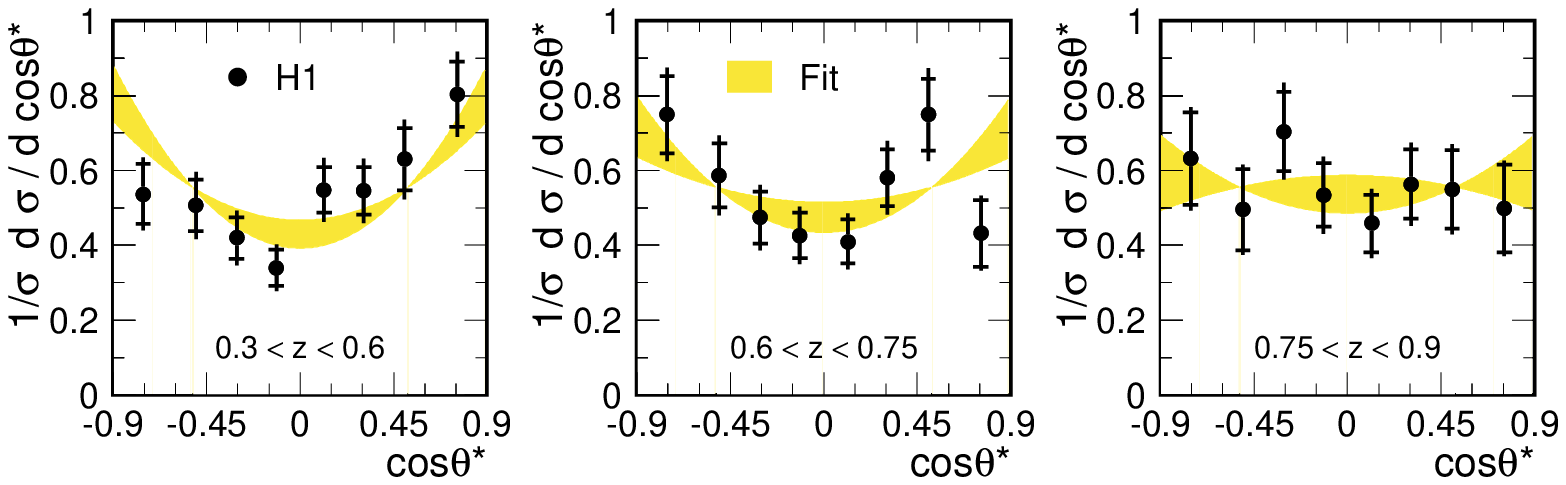,width=\textwidth}
%}
\end{picture}
\caption{\it Differential cross sections 
$1/\sigma\,d\sigma / d\cos\theta^\ast$  in the rest system of the \jpsi\ meson 
for different $z$ regions
 (kinematic region $60<W<240\,\mbox{~GeV}$, $\ptt>1\,\gevt$) normalised for $|\cos\theta^\ast|<0.9$.  
The inner error bars indicate the statistical  uncertainty, while
  the outer error bars include the statistical and  systematic uncertainties added
  in quadrature. The shaded bands reflect fits to the form $\sim
1+\lambda\cos^2{\theta^*}$. The contours of $\pm 1$ standard
deviation are shown.}
 \label{polar1}
%\end{figure}

%\begin{figure}[htbp]
\unitlength1.0cm
\begin{picture}(16,12.3)
%\put(7,13.){\Large Medium $z$}
\put(1.0,-0.8){\epsfig{file=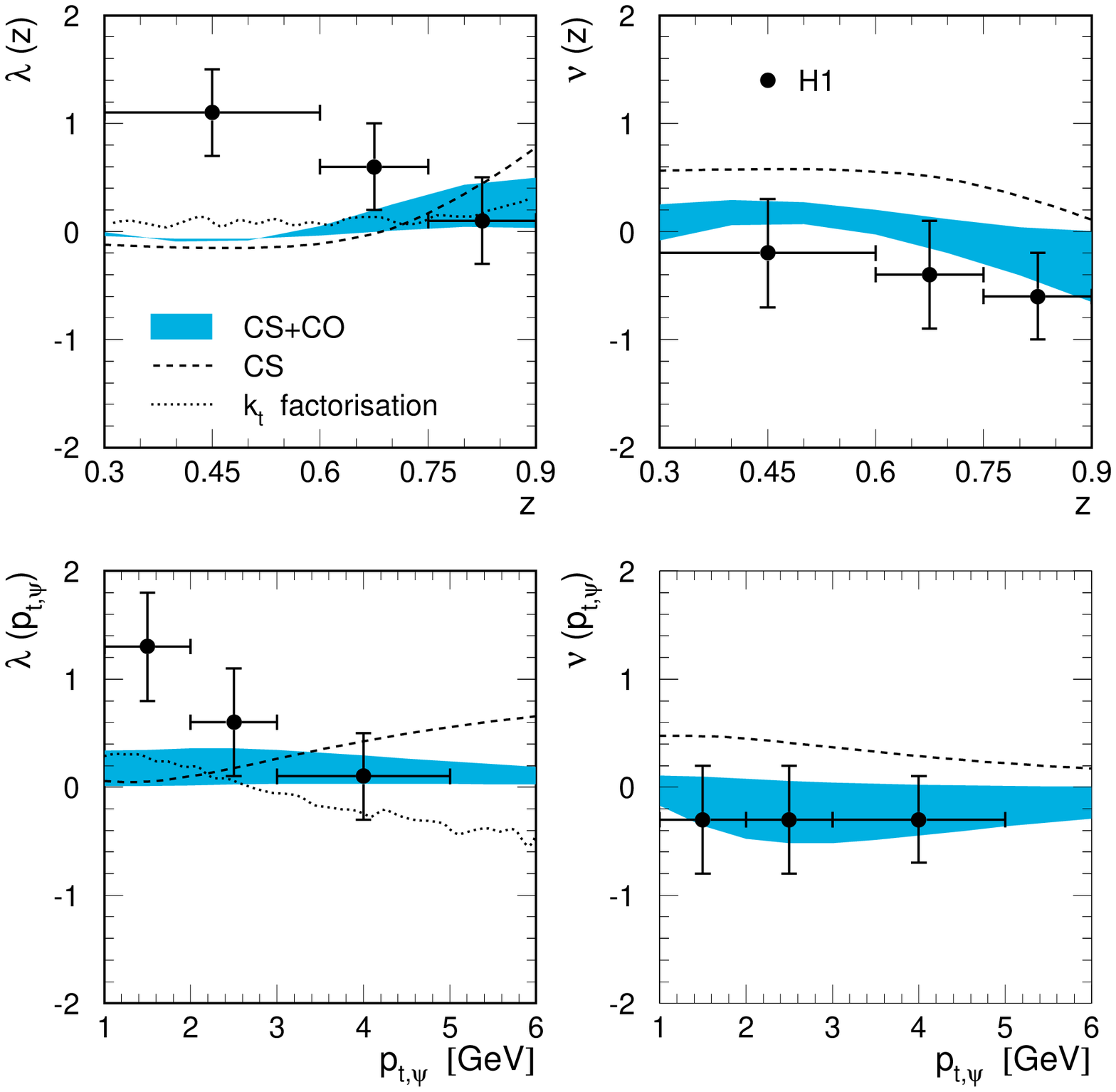,width=14cm}}
\put(7.,11.) {a)}
\put(13.,11.) {b)}
\put(7.,4.5) {c)}
\put(13.,4.5) {d)}
\end{picture}
\caption{\it Polarisation parameters $\lambda$ (a and c) and $\nu$ (b
and d)
in the target frame as functions of $z$ (a and b) and \ptpsi\ (c and d). 
The error bars on the data points correspond to the 
total experimental error. The theoretical
calculations shown are from the NRQCD 
approach \cite{beneke} (shaded bands) with \coloct\ and \colsing\ contributions, while the dashed 
curves show the result from the \colsing\ contribution separately.
For $\lambda$ a calculation within the $k_t$ factorisation 
approach \cite{baranov} is also shown (dotted line).}
\label{polar2}
\end{figure}

The corrected data in the medium $z$ range (dataset~I) are normalised
to the integrated cross sections and fitted to the forms \cite{beneke}:
\begin{eqnarray}
\frac{1}{\sigma}\frac{d\sigma}{d\cos{\theta^*}}&\propto&1+\lambda\cos^2{\theta^*}\, ;
\label{eq:th}\\[.3em]
\frac{1}{\sigma}\frac{d\sigma}{d\Phi^*}&\propto&1+\frac{\lambda}{3}+\frac{\nu}{3}\,\cos{2\Phi^*}\, .\label{eq:ph}
\end{eqnarray} 
The parameters $\lambda$ and $\nu$ can be related to the polarisation of the \jpsi\ 
meson. The cases $\lambda=1$ and $-1$  correspond to transverse and longitudinal 
polarisation of the 
\jpsi\ meson, respectively. Fits of the data to equations (\ref{eq:th})
and (\ref{eq:ph}) are performed in 3 bins of $z$ or in 3 bins of \ptpsi.
The polar angular distribution in $z$ bins is shown as
an example in Fig.~\ref{polar1}. The results for $\lambda$ are listed in table
\ref{tab:polar1} and plotted in Fig.~\ref{polar2}a and c.
Positive values are preferred although a decrease %(2 standard deviations) 
is observed with increasing $z$ and increasing \ptpsi.
The fit of the $\Phi^*$ distribution is also performed in $z$ and \ptpsi\ bins\ using 
the fitted values 
for $\lambda$. The resulting $\nu$ values are slightly negative (see Table~\ref{tab:polar1} and
Fig.~\ref{polar2}b and d) and within errors independent of $z$ and \ptpsi.

\begin{table}\centering
\begin{tabular}{|c|c|c|}
\hline
$z$ interval & $\lambda$ & $\nu$ \\ \hline
   $0.3 - 0.6$  & $1.1 \pm 0.4$ & $-0.2 \pm 0.5$ \\
   $0.6 - 0.75$ & $0.6 \pm 0.4$ & $-0.4 \pm 0.5$ \\
  $0.75 - 0.9$  & $0.1 \pm 0.4$ & $-0.6 \pm 0.4$ \\ \hline \hline
%  1.081160      .4369390      .3670630
%  .6407380      .3911590      .3306020    
%  .1326780      .3879070      .3228980    
% -.2427440      .4459650      .4459360      .2510750E-01  .3053860E-01
% -.4079400      .4326800      .4327940      .4132630E-01  .4984960E-01
% -.5998640      .4202920      .4202740      .6770720E-01  .8305060E-01
\ptpsi\ interval\,[$\gev$] & $\lambda$ & $\nu$ \\ \hline
  $ 1 - 2$ &   $1.3 \pm  0.5$ & $-0.3 \pm 0.5 $ \\ 
  $ 2 - 3$ &   $0.6 \pm  0.5$ & $-0.3 \pm 0.5 $ \\	
  $ >3$     &   $0.1 \pm  0.4$ & $-0.3 \pm 0.4 $ \\	
%  1.342970      .5512660	.4472770
%  .5572250      .4748690      .3877600    
%  .1192930      .4023930      .3319890    
% -.3049820      .4631020      .4631000      .3125790E-01  .3860540E-01
% -.2575170      .4721250      .4721300      .2809350E-01  .3437830E-01
% -.3271000      .3986540      .3986020      .3487450E-01  .4223980E-01  
\hline
\end{tabular}
\caption{\it Fit results for the polarisation parameters $\lambda$ and
$\nu$ as functions of $z$ and \ptpsi. The errors are due to the total
experimental uncertainties.}
  \label{tab:polar1}
\end{table}

In Fig.~\ref{polar2} the results for three LO calculations are shown together with 
the data: the NRQCD prediction including \coloct\ and \colsing\ 
contributions \cite{beneke}, the \colsing\ contribution alone and a
calculation using a $k_t$ factorisation approach and off-shell gluons~\cite{baranov}. 
The errors in the data preclude any firm conclusions. 
None of the three calculations predicts a decrease of $\lambda$ with increasing $z$, 
while a decrease with increasing \ptpsi\ as observed in the data is expected for 
the $k_t$ factorisation ansatz. The full NRQCD prediction
 is also compatible with this. In order to distinguish between full NRQCD and the
\colsing\ contribution alone, measurements at larger \ptpsi\ are required. 
The measured values of $\nu$, for which no prediction is available from 
the $k_t$ factorisation approach, favour the full NRQCD prediction.
%, but is within the large errors also consistent with the \colsing\ contribution alone. 

%----------------------------------------------------------------------

\section{Summary and Conclusions}
An analysis of inelastic photoproduction of \jpsi\ mesons is presented. 
The kinematic region covers 
$60<W_{\gamma p}<260\,{\rm GeV}$, \ptt$>1\,\gevt$ and
$0.05<z<0.9$. Cross sections in the low $z$ region, $z\lsim0.3$, are presented 
for the first time.
The data can be described by boson gluon fusion calculations. In the
low $z$ range the agreement between data and (LO) calculations is
improved by including resolved photon processes although the uncertainties due to
contributions from $b$ decays are appreciable in this region.
The differential cross sections are compared with calculations in three 
different theoretical frameworks.

Firstly, in the medium $z$ range, $0.3\leq z\leq0.9$, (double) differential cross sections 
are compared with calculations in the \csm\ (CSM) in NLO for direct photons. 
These  are found to give a good description of the distributions 
in \wgp, $z$ and \ptt\ (tested up to a value of \ptt$\sim60\,\gevt$). 
The NLO calculations show a hard \ptr\ spectrum which describes the
data well in contrast to the LO calculation. The distribution of the
transverse momentum of the \jpsi\ is also found to be well described
in three separate regions of $z$. 
The estimated theoretical uncertainties due to the uncertainties in the charm mass and 
$\alpha_s$ are much larger than the experimental errors,
so the data may be used to constrain parameters of the model.

A second comparison in this medium $z$ range is made 
with a LO calculation within the CSM, 
allowing the interacting gluons to have transverse momentum 
($k_t$ factorisation approach). This Monte Carlo calculation  
is also found to give a good description of the data.

The third comparison in the whole $z$ range ($0.05<z<0.9$) involves LO non-relativistic QCD
calculations (NRQCD) including \coloct\
as well as \colsing\ contributions. This is of importance since 
the \csm\ (in LO) does not describe charmonium production in 
\ppbar\ collisions and has fundamental theoretical problems in the 
description of $p$-wave states. 
The  present differential cross sections can be reasonably well described 
in shape by the NRQCD calculations when including direct
and resolved contributions. The normalisation depends on the
details of the calculations, in particular on the chosen octet long
distance matrix elements (LDMEs). 
The present photoproduction data clearly favour very low values of the octet LDMEs.
At high $z$, resummed NRQCD calculations applicable at large $\ptpsi$, tend to improve the agreement with the data. 
At low $z$ the inclusion of \coloct\ contributions 
improves the agreement between data and theory. 
However, decays of $b$ flavoured hadrons are expected to play an important role 
here and are neglected in the present analysis.
 
The \ptt\ distribution is reasonably reproduced by the NRQCD calculation, 
which however gives a slightly softer \ptt\ dependence 
than that measured in the medium and in the low $z$ ranges. Including NLO 
effects would probably improve the agreement with the measurements as
already seen in the CSM. 

The polarisation parameters of the \jpsi\ meson have been measured as
a function of \ptpsi\ and $z$. Within present experimental and
theoretical errors  NRQCD, the CSM and the 
$k_t$ factorisation approach all fit the data reasonably well.

With an appropriate choice of parameters,
theoretical calculations within the NRQCD approach, constrained by
results from \ppbar\ collisions can describe
the present photoproduction measurements. Next-to-leading order corrections 
are likely to improve the agreement even after \coloct\ contributions are included.
The data have the potential to reduce the current large
uncertainties in the phenomenological parameters. This contributes to the development 
of a unified understanding of charmonium production in     
different environments such as \ppbar, $ep$ and $\gamma p$ collisions.     

%%%%%%%%%%%%%%%%%%%%%%%%%%%%%%%%%%%%%%%%%%%%%%%%%%%%%%%%%%%%
\section*{Acknowledgements}

We are grateful to the HERA machine group whose outstanding
efforts have made and continue to make this experiment possible. 
We thank
the engineers and technicians for their work in constructing and now
maintaining the H1 detector, our funding agencies for 
financial support, the
DESY technical staff for continual assistance, 
and the DESY directorate for the
hospitality which they extend to the non DESY 
members of the collaboration. We want to thank 
M. Kr\"amer for close collaboration and S. Wolf, B.A. Kniehl
and S.P. Baranov for many discussions and for making their theoretical 
calculations available to us.
%-------------------------------------------------------------------------------
%-------------------------------------------------------------------------------
\clearpage

\clearpage

\begin{table}\centering
\begin{tabular}{|l|l|}
\hline
Parameters NLO CSM (M. Kr\"amer \cite{csm_nlo})& Figures \ref{fig1} and \ref{double} \\
\hline
PDF & MRST\cite{mrst}\\
Renormalisation/Factorisation scale& $\sqrt{2}\,m_c$, max$[\sqrt{2}\,m_c, 1/2\,\sqrt{m_c^2 + p_t^2}]$ (for \ptt)\\
$\langle{\cal O}[1,^3\!S_1]\rangle$ & $1.16\,\gev^{3}$\\
$m_c$ & $1.3\leq m_c\leq1.5\,\gev$\\
$\alpha_s(M_Z)$& $0.1200\pm0.0025$\\
\hline\hline
Parameters NRQCD (M. Kr\"amer \cite{kraemer})& Figures \ref{cslowz2} and \ref{allz}\\
\hline
PDF & GRV(LO)\cite{GRV} for proton and photon\\
$\Lambda^{(4)}_{LO}$&$200\,\mev$\\
Renormalisation/Factorisation scale& $2\,m_c$\\
$m_c$ & $1.5\,\gev$\\
$\langle{\cal O}[1,^3\!S_1]\rangle$ & $1.16\,\gev^{3}$\\
$\langle{\cal O}[8,^3\!S_1]\rangle$ & $(0.3-2.0) \cdot 10^{-2}\gev^{3}$\\
$\langle{\cal O}[8,^1\!S_0]\rangle+3.5\,\langle{\cal O}[8,^3\!P_0]\rangle/m_c^2$ & $(1.0-10) \cdot 10^{-2}\gev^{3}$\\
\hline\hline
Resummed calculations (Beneke et al. \cite{wolf})&Figure~\ref{wolf}\\
\hline
$\langle{\cal O}[8,^1\!S_0]\rangle+3.1\,\langle{\cal O}[8,^3\!P_0]\rangle/m_c^2$ &$ 1.5 \cdot 10^{-2}\gev^{3}$\\
$\langle{\cal O}[8,^3\!S_1]\rangle $&$ (0.5-1.0) \cdot 10^{-2}\gev^{3}$\\
other parameters as in the NRQCD calculation \cite{kraemer}&\\
\hline\hline
Parameters NRQCD (Kniehl et al. \cite{kniehl}) & Figure~\ref{allz}\\
\hline
PDF & CTEQ5M \cite{cteq5}\,/\,GRV-HO for proton and photon\\
$\Lambda^{(4)}_{\overline{MS}}$&$326\,\mev$\\
Renormalisation/Factorisation scale& $\sqrt{4\,m_c^2+p_t^2}$ \\
$m_c$ & $M_\psi/2$\\
$\langle{\cal O}[1,^3\!S_1]\rangle$ & $(1.3\pm0.09)\,\gev^{3}$\\
$\langle{\cal O}[8,^1\!S_0]\rangle+3.54\, \langle{\cal O}[8,^3\!P_0]\rangle/m_c^2$ &$ (0.572\pm0.184) \cdot 10^{-2}\gev^{3}$\\
$\langle{\cal O}[8,^3\!S_1]\rangle$ & $(0.273\pm0.045) \cdot 10^{-2}\gev^{3}$\\
\hline
% Kraemer (review):
%$O[8,^1\!S_0]\rangle = 1.2 \cdot 10^{-2}GeV^{3}$
%$O[8,^3\!P_0)/m_c^2 = 1.2 \cdot 10^{-2}GeV^{3}$
%\rightarrow  M_{3.1}=4.9 \cdot 10^{-2}GeV^{3}$
\end{tabular}
\caption{\it Parameters used in the QCD calculations which are
compared to the data. The expressions ``$\langle{\cal O}[8,^3\!S_1]\rangle$'' etc. denote
the long distance matrix elements. The \coloct\ LDMEs are extracted
from high $p_t$ \jpsi\ production in \ppbar\ collisions for the NRQCD 
calculations of \cite{kniehl,kraemer} and from data on $B$
decays in \cite{wolf}. For more details and for a discussion of the  
uncertainties see these references.   }
  \label{tab:qcd}
\end{table}

\begin{table}[h]
 \begin{center}\begin{footnotesize}
 \begin{tabular}{|r|r|c|c|}
 \hline
 \multicolumn{1}{|c|}{$W_{\gamma p}\ [{\rm GeV}]$} & \multicolumn{1}{c|}{$\langle W_{\gamma p} \rangle$} &
 \multicolumn{2}{c|}{$\sigma_{\gamma p}\ [{\rm nb}]$} \\
  & & $0.3<z<0.9$ & $0.3<z<0.8$ \\ \hline \hline
$ 60 -  80$&$ 69.3$&$ 23.0 \pm   2.5 \pm   3.2$ & $ 17.5 \pm   2.0 \pm   2.4$\\ \hline
$ 80 - 100$&$ 89.4$&$ 29.7 \pm   2.2 \pm   4.1$ & $ 21.4 \pm   1.7 \pm   3.0$\\ \hline
$100 - 120$&$109.5$&$ 34.1 \pm   2.3 \pm   4.7$ & $ 27.2 \pm   2.0 \pm   3.8$\\ \hline
$120 - 140$&$129.6$&$ 36.7 \pm   2.6 \pm   5.1$ & $ 27.7 \pm   2.2 \pm   3.9$\\ \hline
$140 - 160$&$149.6$&$ 37.3 \pm   3.0 \pm   5.2$ & $ 27.0 \pm   2.4 \pm   3.8$\\ \hline
$160 - 180$&$169.6$&$ 42.0 \pm   3.9 \pm   5.8$ & $ 31.8 \pm   3.1 \pm   4.4$\\ \hline
$180 - 210$&$194.2$&$ 41.7 \pm   4.9 \pm   5.8$ & $ 31.1 \pm   3.7 \pm   4.3$\\ \hline
 & & \multicolumn{1}{r|}{$ +  5.9$} & \multicolumn{1}{r|}{$ +  4.1$}\\ 
\rb{$210 - 240$}&\rb{$224.2$}&\rb{$ 37.8 \pm   7.8$} $-  5.0$ & \rb{$ 26.4 \pm   5.8$} $-  3.5$\\ \hline
  \end{tabular}\end{footnotesize}
 \end{center}
% \end{table}
 %
% \begin{table}[htbp]
 \begin{center}\begin{footnotesize}
 \begin{tabular}{|r|r|c|c|c|}
 \hline
 \multicolumn{1}{|c|}{$z$} & \multicolumn{1}{c|}{$\langle z \rangle$} &  \multicolumn{3}{c|}{${\rm d}\sigma_{\gamma p}/{\rm d}z
\ [{\rm nb}]$} \\
  & & $p_{t,\psi}>1\,{\rm GeV}$ & $p_{t,\psi}>2\,{\rm GeV}$ & $p_{t,\psi}>3\,{\rm GeV}$ \\ \hline \hline
$ 0.30 -  0.45$&$0.38$ & $ 27.5 \pm   2.4 \pm   3.8$ & $ 11.6 \pm   1.4 \pm   1.6$ & $  4.3 \pm   0.8 \pm   0.6$\\ \hline
$ 0.45 -  0.60$&$0.53$ & $ 51.7 \pm   2.8 \pm   7.2$ & $ 20.8 \pm   1.6 \pm   2.9$ & $  7.3 \pm   0.8 \pm   1.0$\\ \hline
$ 0.60 -  0.75$&$0.68$ & $ 64.3 \pm   3.4 \pm   8.9$ & $ 26.8 \pm   2.0 \pm   3.7$ & $ 10.0 \pm   1.1 \pm   1.4$\\ \hline
$ 0.75 -  0.90$&$0.83$ & $ 77.4 \pm   5.0 \pm  10.8$ & $ 28.3 \pm   2.6 \pm   3.9$ & $ 10.5 \pm   1.3 \pm   1.5$\\ \hline
  \end{tabular}\end{footnotesize}
 \end{center}
% \end{table}
 %
% \begin{table}[htbp]
 \begin{center}\begin{footnotesize}
 \begin{tabular}{|r|r|c|}
 \hline
 \multicolumn{1}{|c|}{$\ptt [{\rm GeV}^2]$} & \multicolumn{1}{c|}{$\langle \ptt \rangle$} &
 ${\rm d}\sigma_{\gamma p}/{\rm d}\ptt\ [{\rm nb/GeV}^2]$ \\ \hline \hline
$ 1.0 -  2.1$&$ 1.54$&$   9.45 \pm    0.61 \pm    1.31$\\ \hline
$ 2.1 -  3.5$&$ 2.78$&$   5.34 \pm    0.37 \pm    0.74$\\ \hline
$ 3.5 -  5.4$&$ 4.38$&$   2.77 \pm    0.22 \pm    0.38$\\ \hline
$ 5.4 -  7.6$&$ 6.43$&$   1.46 \pm    0.13 \pm    0.20$\\ \hline
$ 7.6 - 10.0$&$ 8.74$&$  0.887 \pm   0.088 \pm   0.123$\\ \hline
$10.0 - 13.5$&$11.6$&$  0.452 \pm   0.053 \pm   0.063$\\ \hline
$13.5 - 20.0$&$16.4$&$  0.160 \pm   0.022 \pm   0.022$\\ \hline
$20.0 - 26.5$&$23.0$&$ 0.0815 \pm  0.0120 \pm  0.0113$\\ \hline
$26.5 - 40.0$&$32.3$&$ 0.0236 \pm  0.0051 \pm  0.0033$\\ \hline
$40.0 - 60.0$&$48.5$&$ 0.0051 \pm  0.0016 \pm  0.0007$\\ \hline
  \end{tabular}\end{footnotesize}
 \end{center}
  \caption{\it  Medium $z$: Cross sections in bins of  $W_{\gamma p}$,
 $z$ and \ptt\ ($0.3<z<0.9$, 
$60<\wgp<240\,\gev$).}
 \label{tab_medz_xsec_pt}
% \end{table}
 %
% \begin{table}[htbp]
\vspace{-0.5cm}
 \begin{center}\begin{footnotesize}
 \begin{tabular}{|r|r|c|c|c|}
 \hline
 \multicolumn{1}{|c|}{$\ptt [{\rm GeV}^2]$} & \multicolumn{1}{c|}{$\langle \ptt \rangle$} &
 \multicolumn{3}{c|}{$ {\rm d}^2\sigma_{\gamma p}/{\rm d}\ptt {\rm d}z\ [{\rm nb/GeV}^2]$} \\
  & & $0.3<z<0.6$ & $0.6<z<0.75$ & $0.75<z<0.9$ \\ \hline \hline
$ 1.0 -  2.1$&$  1.54$&$   12.2 \pm     1.2 \pm     1.7$&$   14.0 \pm     1.7 \pm     1.9$&$   22.6 \pm     2.8 \pm     3.1$\\ \hline
$ 2.1 -  3.5$&$  2.78$&$   5.93 \pm     0.65 \pm    0.82$&$  11.9 \pm    1.33 \pm    1.66$&$  11.4 \pm    1.8 \pm    1.6$\\ \hline
$ 3.5 -  5.4$&$  4.38$&$   3.11 \pm     0.41 \pm    0.43$&$   6.14 \pm    0.77 \pm    0.85$&$   6.27 \pm    1.03 \pm    0.87$\\ \hline
$ 5.4 -  7.6$&$  6.43$&$   2.08 \pm     0.26 \pm    0.29$&$   2.61 \pm    0.46 \pm    0.36$&$   2.72 \pm    0.57 \pm    0.38$\\ \hline
$ 7.6 - 10.0$&$  8.74$&$   1.17 \pm     0.18 \pm    0.16$&$   1.56 \pm    0.27 \pm    0.22$&$   2.57 \pm    0.54 \pm    0.36$\\ \hline
$10.0 - 13.5$&$ 11.6$&$  0.530 \pm   0.103 \pm   0.074$&$  1.14 \pm   0.208 \pm   0.158$&$  1.01 \pm   0.24 \pm   0.14$\\ \hline
$13.5 - 20.0$&$ 16.4$&$  0.184 \pm   0.039 \pm   0.026$&$  0.547 \pm   0.113 \pm   0.076$&$  0.276 \pm   0.083 \pm   0.038$\\ \hline
$20.0 - 26.5$&$ 23.0$&$  0.145 \pm   0.029 \pm   0.020$&$  0.143 \pm   0.040 \pm   0.020$&$  0.110 \pm   0.039 \pm   0.015$\\ \hline
$26.5 - 40.0$&$ 32.3$&$ 0.0323 \pm  0.0097 \pm  0.0045$&$ 0.0296 \pm  0.0169 \pm  0.0041$&$ 0.0804 \pm  0.0284 \pm  0.0112$\\ \hline
$40.0 - 60.0$&$ 48.5$&$ 0.0113 \pm  0.0059 \pm  0.0016$&$ 0.0059 \pm  0.0029 \pm  0.0008$&$ 0.0057 \pm  0.0033 \pm  0.0008$\\ \hline
  \end{tabular}\end{footnotesize}
 \end{center}
 \caption{\it  Medium $z$: Double differential cross sections
  in $z$ and \ptt\ ($60<\wgp<240\,\gev$).}
 \label{tab_medz_xsec_double}
 \end{table}
\begin{table}[t]
  \begin{center}\begin{footnotesize}
 \begin{tabular}{|r|r|c|}
 \hline
 \multicolumn{1}{|c|}{$W_{\gamma p}\ [{\rm GeV}]$} & \multicolumn{1}{c|}{$\langle W_{\gamma p} \rangle$} &
 $\sigma_{\gamma p}\ [{\rm nb}]$ \\ \hline \hline
$120 - 150$&$133.8$&$ 12.1 \pm   1.5 \pm   2.6$\\ \hline
$150 - 180$&$163.9$&$ 12.8 \pm   2.1 \pm   2.5$\\ \hline
$180 - 260$&$213.7$&$ 15.6 \pm   2.5 \pm   3.1$\\ \hline
 \end{tabular}\end{footnotesize}
 \end{center}
%   \caption{\it Low $z$: Cross sections in bins of
%  $W_{\gamma p}$ ($0.05<z<0.45$, $\ptpsi>1\,\gev$)}
% \label{tab_lowz_xsec_w}
% \end{table}
 %
 %
% \begin{table}[htbp]
  \begin{center}\begin{footnotesize}
 \begin{tabular}{|r|r|c|}
 \hline
 \multicolumn{1}{|c|}{$\ptt [{\rm GeV}^2]$} & \multicolumn{1}{c|}{$\langle \ptt \rangle$} &
 ${\rm d}\sigma_{\gamma p}/{\rm d}\ptt\ [{\rm nb/GeV}^2]$ \\ \hline \hline
$ 1.0 -  2.1$&$ 1.54$&$   3.56 \pm    0.83 \pm    0.68$\\ \hline
$ 2.1 -  3.5$&$ 2.78$&$   1.85 \pm    0.42 \pm    0.35$\\ \hline
$ 3.5 -  5.4$&$ 4.38$&$   1.24 \pm    0.30 \pm    0.24$\\ \hline
$ 5.4 -  7.6$&$ 6.43$&$  0.875 \pm   0.177 \pm   0.167$\\ \hline
$ 7.6 - 10.0$&$ 8.75$&$  0.489 \pm   0.115 \pm   0.095$\\ \hline
$10.0 - 13.5$&$11.6$&$  0.267 \pm   0.068 \pm   0.053$\\ \hline
$13.5 - 20.0$&$16.4$&$  0.117 \pm   0.030 \pm   0.024$\\ \hline
$20.0 - 40.0$&$27.8$&$ 0.0197 \pm  0.0044 \pm  0.0043$\\ \hline
 \end{tabular}\end{footnotesize}
 \end{center}
 \caption{\it  Low $z$: Differential cross sections in bins of  $W_{\gamma p}$
 and \ptt\ ($0.05<z<0.45$, $120<\wgp<260\,\gev$).}
 \label{tab_lowz_xsec_pt}
% \end{table}

%\begin{table}[h!]
  \begin{center}\begin{footnotesize}
 \begin{tabular}{|r|r|c|}
 \hline
 \multicolumn{1}{|c|}{$z$} & \multicolumn{1}{c|}{$\langle z \rangle$} &  \multicolumn{1}{c|}{${\rm d}\sigma_{\gamma p}/{\rm d}z
\ [{\rm nb}]$} \\ \hline \hline
$ 0.05 -  0.15$&$0.10$ & $ 35.1 \pm   8.3 \pm   8.5$\\ \hline
$ 0.15 -  0.30$&$0.23$ & $ 31.1 \pm   4.1 \pm   6.2$\\ \hline
$ 0.30 -  0.45$&$0.38$ & $ 36.7 \pm   4.0 \pm   7.0$\\ \hline
 \hline
$ 0.30 -  0.45$&$0.38$ & $ 31.8 \pm   3.6 \pm   4.4$\\ \hline
$ 0.45 -  0.60$&$0.53$ & $ 58.7 \pm   4.3 \pm   8.2$\\ \hline
$ 0.60 -  0.75$&$0.68$ & $ 80.2 \pm   5.3 \pm  11.2$\\ \hline
$ 0.75 -  0.90$&$0.83$ & $ 98.7 \pm   7.5 \pm  13.7$\\ \hline
 \end{tabular}\end{footnotesize}
 \end{center}
 \caption{\it Total $z$ range: Differential cross sections in bins of
 $z$ ($120<\wgp<260\,\gev$ and $\ptpsi>1\,\gev$). The points at
 $z=0.375$ are statistically correlated, the first one is obtained
 from dataset~II the second one from dataset~III.}
 \label{tab_allz_xsec}
 \end{table}
\vfill
\end{document}